\documentclass[11pt,epsf]{article}

\usepackage{graphicx}
\usepackage{amsfonts,amssymb,amsmath}
\usepackage{subfigure}
\usepackage{color}

\begin{document}
\begin{titlepage}

\begin{center}
{\Large\bf Instability of Black Holes with a Gauss-Bonnet Term}\\
\vspace{.4in}

  {$\rm{Wha-Keun \,\, Ahn}^{\S}$}\footnote{\it email: wkahn@sogang.ac.kr}\, \,
  {$\rm{Bogeun \,\, Gwak}^{\S}$}\footnote{\it email: rasenis@sogang.ac.kr}\, \,
  {$\rm{Bum-Hoon \,\, Lee}^{\S\dag *}$}\footnote{\it email: bhl@sogang.ac.kr}\,\,
  {$\rm{Wonwoo \,\, Lee}^{\S}$}\footnote{\it email: warrior@sogang.ac.kr} \\

  {\small \S \it Center for Quantum Spacetime, Sogang University, Seoul 121-742, Korea}\\
  {\small \dag \it Department of Physics, Sogang University, Seoul 121-742, Korea}\\
  {\small * \it Asia Pacific Center for Theoretical Physics, Pohang 790-784, Korea}\\

\vspace{.5in}
\end{center}

{\abstract
{We investigate the fragmentation instability of hairy black holes in the theory with a Gauss$-$Bonnet(GB) term in asymptotically flat spacetime. Our approach is through the non-perturbative fragmentation instability. By this approach, we investigate whether the initial black hole can be broken into two black holes by comparing the entropy of the initial  black hole with the sum of those of two fragmented black holes. The relation between the black hole instability and the GB coupling with dilaton hair are presented. We describe the phase diagrams with respect to the mass of the black hole solutions and coupling constants. We find that a perturbatively stable black hole can be unstable under fragmentation.}}

\end{titlepage}

\newpage

\section{Introduction \label{sec1}}

The no-hair theorem states that the black holes in Einstein$-$Maxwell theory are characterized by only there mass, electric charge, and angular momentum\,\cite{Bekenstein:1972ky}\,. All other observable parameters as regards a black hole are hidden in the event horizon, i.e.\ the contributions from other parameters cannot be accessible to an outside observer. Various gravity theories motivated by string theory and cosmology have received more and more attention. In this perspective, various kinds of black holes with different hairs have been investigated \cite{drhs}. The complex dilaton nonminimally coupled with the Maxwell field presents the first hairy black hole\,\cite{Gibbons:1987ps}\,. The black hole hairs were categorized into two types, primary and secondary. A primary hair independently gives a new quantum number to a black hole, so the black hole states are expanded\,\cite{Coleman:1991ku}\,. On the other hand, a secondary hair is determined by the primary hair\,\cite{Coleman:1991jf}\,. The black hole dilaton hair is classified as secondary hair, because the dilaton field appears to be coupled to a Maxwell field. Recently, dilaton hairs have been discovered in many other theories of gravitation. One theory motivated to show the next-leading order effect of the inverse string tension $\alpha'$\, ($16\alpha\kappa$ in the present paper) includes higher-order curvature called the Gauss$-$Bonnet(GB) term\,\cite{Boulware:1986dr}\,. The GB term is the simplest one in the low energy effective supergravity action. In four dimensions, the presence of a GB term does not have any ghost particles or any problem of unitarity. In addition, the GB term does not change the second-order equation of motion \cite{Boulware:1986dr,Callan:1985ia}\,.

In the cosmological model, the dilatonic Einstein$-$Gauss$-$Bonnet(DEGB) theory can provide the possibility of avoiding the initial singularity of the universe \cite{Antoniadis:1992rq}. It may violate the energy condition in the singularity theorem thanks to the presence of that term. Recently, the specific inflationary model with the GB term has been studied \cite{Guo:2010jr}\,. In the DEGB theory, the nontrivial real dilaton field appears in the black hole solution\,\cite{Guo:2008hf,Guo:2008hf2,Kanti:1995vq,Torii:1996yi,Kanti:1997br,Torii:1998gm}\, as a scalar hair. The black hole hair is secondary\,\cite{Kanti:1995vq}\, in the DEGB theory, because the scalar hair is determined by the mass of the black hole. There exists a minimum black hole mass, below which black hole solutions do not exist. Above that minimum mass there exist upper and lower branch solutions. The upper branch solutions are stable under linear perturbations and approach the Schwarzschild black holes in the large mass limit. Depending on the couplings, the lower branch solutions are unstable under linear perturbations, and they end at a singular solution\,\cite{Kanti:1997br,Torii:1998gm,Moura:2012fq}\,. Our goal is to investigate black hole instability by a non-perturbative method on the upper branch.

In higher-dimensional spacetime, there exist various rotating black holes for given angular momentum. The Myers$-$Perry(MP) black hole is a Kerr black hole generalized to higher dimensions\,\cite{Myers:1986un}\,. The black ring is another type of solutions, which becomes more stable than MP black hole in higher angular momentum\,\cite{Emparan:2008eg,Emparan:2007wm,Emparan:2003wm,Gwak:2011rp}\,. For large angular momentum, a black hole can undergo fragmentation\,\cite{Emparan:2003wm}\,. Fragmentation is based on the entropy\,\cite{Bekenstein:1973ur}\, preference between the solutions. Fragmentation allows for the upper or lower bound of black hole charges\,\cite{Gwak:2014xra}\,. Unstable black holes are important and related to the non-equilibrium states in the anti-de Sitter/conformal field theory(AdS/CFT) correspondence\,\cite{Bai:2012ci}\,.

In this paper, we compute and describe the fragmentation instabilities of the black hole with a GB term arising in asymptotically flat 4-dimensional spacetime in which a fragmentation instability have not been done before. We found that a stable solution under perturbation can be unstable under the fragmentation. We show the black hole instability depending on the GB couplings. We also present the phase diagrams on parameter space.

The outline of this paper is as follows: in Sect.\ \ref{sec2}\,, we introduce our basic framework and numerical construction of the black holes for the theory where the dilaton field is coupled with the GB term. We numerically solve the equations of motions to construct hairy black holes. We explain the black hole properties in the theory. In Sect.\ \ref{sec3}\,, we describe instabilities of black holes. In Sect.\ \ref{sec4}\,, we numerically investigate black hole instabilities through fragmentation. Black hole phase diagrams are presented in parameter space. In Sect.\ \ref{sec5}\,, we summarize and discuss our results.

\section{Hairy black holes in DEGB theory \label{sec2}}

As the simplest model of the effective low energy supergravity action, for the gravity theory we are motivated to use a GB term to investigate the next-leading order effect of the inverse string tension. The GB term is a good model to show the quantum effect. This effect should affect instability of black hole with GB term. We are interested in the instability of a black hole due to fragmentation. The fragmentation phenomena of a black hole may occur by a large quantum or thermal phase transition. Einstein gravity does not allow for these phenomena. In this perspective, one could introduce the Einstein theory of gravity with a GB term as the effective theory including a quantum correction.

\subsection{Action and black hole solutions}

To explore the fragmentation phenomena, we consider the action as follows:
\begin{equation}
I= - \int_{\mathcal M} \sqrt{-g} d^4 x \left[ \frac{R}{2\kappa}
-\frac{1}{2}{\nabla_\alpha}\Phi {\nabla^\alpha}\Phi
 +\alpha e^{-\gamma\Phi} R_{\rm GB}^2 \right] + \oint_{\partial \mathcal M} \sqrt{-h} d^3 x
\frac{K-K_o}{\kappa}\,, \label{gb-action}
\end{equation}
where $g=\det g_{\mu\nu}\,$, $\kappa \equiv 8\pi G\,$, and $R\,$ denotes the scalar curvature of the spacetime $\mathcal M$\,. The higher-curvature GB term is given by $R^{2}_{\rm GB} = R^2 - 4 R_{\mu\nu}R^{\mu\nu} +  R_{\mu\nu\rho\sigma} R^{\mu\nu\rho\sigma}$\,. The action has a dilaton field $\Phi\,$ coupled with the GB term $\alpha e^{-\gamma\Phi}\,$ where $\alpha\,$ and $\gamma\,$ are constants, which we will call as dilatonic Einstein$-$Gauss$-$Bonnet(DEGB) theory.  The second term on the right-hand side is the boundary term \cite{York, giha} in which $h$ is the determinant of the first fundamental form, $K$ and $K_o$ are the traces of the second fundamental form of the boundary $\partial \mathcal M$ for the metric $g_{\mu\nu}$ and $\eta_{\mu\nu}\,$, respectively. The gravitational field equations can be obtained properly from a variational principle with this boundary term. We adopt the sign conventions in Ref.\ \cite{misner}. The action Eq.~(\ref{gb-action}) is symmetric under
\begin{eqnarray}
\gamma\rightarrow -\gamma,\,\,\,\,\Phi\rightarrow -\Phi\,.
\end{eqnarray}
This allows for positive $\gamma$ values without loss of generality. One can eliminate the coupling $\alpha$ dependency by a $r\rightarrow\frac{r}{\sqrt{\alpha}}$ transformation \cite{Guo:2008hf2}. Under the transformation, the action Eq.~(\ref{gb-action}) corresponds to the $\alpha=1$ case. Non-zero $\alpha$ coupling cases can be generated by $\alpha$ scaling, but the behaviors for the $\alpha=0$ case  cannot be generated in this way. To show a continuous change to $\alpha=0$, we keep the parameter $\alpha$ in the action.

For DEGB theory with given non-zero couplings $\alpha$ and $\gamma$\,, one can see DEGB black holes with a hair. There does not exist black hole solutions without a hair in DEGB theory. If we have $\Phi=0$ exists in DEGB theory, dilaton equation of motion in Eq.~(\ref{tmn}) has only $R^2_{GB}$ term. However, the GB term should be non-zero, so it cannot be satisfy in Eq.~(\ref{tmn}). The coupling $\alpha$ could be absorbed in the redefinition of $r$ as in Ref.~\cite{Guo:2008hf2}, where the black hole properties depend on $\alpha$ scale $\alpha$ except for $\alpha=0$. For the coupling $\alpha=0$, the solutions become a Schwarzschild black hole in Einstein gravity, and $\alpha$ is not absorbed into the radial coordinate $r$ in this work.

Setting $\gamma=0$, the DEGB theory becomes the Einstein$-$Gauss$-$Bonnet(EGB) theory. The EGB black hole solution with a single coupling $\alpha$ is the same as the Schwarzschild one. This is because the GB term does not contribute to the equations of motion. However, the GB term contributes to the black hole entropy and influences the stability.

From the action (\ref{gb-action}), we obtain the Einstein equations and the scalar field equation,
\begin{equation}
R_{\mu\nu} - \frac{1}{2}g_{\mu\nu}R  = \kappa \left( \partial_{\mu}
\Phi \partial_{\nu}\Phi -\frac{1}{2}g_{\mu\nu}
\partial_{\rho} \Phi \partial^{\rho} \Phi  + T_{\mu\nu}^{GB}
\right)\,,
\end{equation}
\begin{equation}
\frac{1}{\sqrt{-g}} \partial_{\mu} [\sqrt{-g}
g^{\mu\nu}\partial_{\nu}\Phi] - \alpha\gamma e^{-\gamma\Phi} R_{GB}^2 =0\,, \label{tmn}
\end{equation}
where the GB term contributes to the energy-momentum tensor
\begin{eqnarray}
T_{\mu\nu}^{GB} &=& -8 \alpha (\nabla^{\rho}\nabla^{\sigma} e^{-\gamma\Phi} R_{\mu\rho\nu\sigma}
-\square e^{-\gamma\Phi} R_{\mu\nu} + 2\nabla_{\rho} \nabla_{(\mu} e^{-\gamma\Phi} {R^{\rho}}_{\nu)}
-\frac{1}{2} \nabla_{\mu}\nabla_{\nu} e^{-\gamma\Phi} R)
\nonumber \\
&&+4\alpha(2\nabla_{\rho}\nabla_{\sigma}  e^{-\gamma\Phi}
R^{\rho\sigma} - \square e^{-\gamma\Phi} R) g_{\mu\nu}\,,
\end{eqnarray}
and $\square \equiv \nabla_{\mu} \nabla^{\mu}$\, is the d'Alembertian.

In this section, we follow the procedure of Ref.\ \cite{Kanti:1995vq}. We consider a spherically symmetric static spacetime with the metric
\begin{equation}
ds^2 = -e^{X(r)}dt^2 + e^{Y(r)} dr^2 + r^2(d\theta^2 + \sin^2\theta d\varphi^2)\,, \label{metric}
\end{equation}
where the metric functions depend only on $r$. Then the dilaton field equation turns out to be
\begin{eqnarray}
\Phi'' + \Phi' \left( \frac{X'-Y'}{2} +\frac{2}{r} \right) &=& - \frac{4\alpha\gamma e^{-\gamma\Phi}}{r^2} \left[ X'Y'e^{-Y}+ (1-e^{-Y})\left( X'' + \frac{X'}{2} (X'-Y') \right)\right]\,,
    \label{feom}
\end{eqnarray}
Also, there are three Einstein equations for $(tt)$, $(rr)$, and $(\theta\theta)$ components, as follows:
\begin{eqnarray}
Y' \left(1- \frac{4\alpha\gamma\kappa e^{-\gamma\Phi}\Phi'}{r} (1-3e^{-Y}) \right)  =  \frac{\kappa r \Phi'^2}{2} + \frac{1-e^Y}{r} -\frac{8\alpha\gamma\kappa e^{-\gamma\Phi}}{r} (\Phi''-\gamma\Phi'^2)(1-e^{-Y})\,, \label{eeom1}
\end{eqnarray}
\begin{equation}
X'\left(1-\frac{4\alpha\gamma\kappa e^{-\gamma\Phi}\Phi'}{r} (1-3e^{-Y}) \right) = \frac{\kappa r\Phi'^2}{2} + \frac{(e^Y -1)}{r}\,,    \label{eeom2}
\end{equation}
\begin{eqnarray}
X'' &+ &\left( \frac{X'}{2} +\frac{1}{r} \right) (X'-Y')\nonumber \label{eeom3}\\
&&=-\kappa\Phi'^2 - \frac{8\alpha\gamma\kappa e^{-\gamma\Phi-Y}}{r} \left( \Phi'X'' + (\Phi''-\gamma\Phi'^2)X' + \frac{\Phi'X'}{2} (X'-3Y') \right),
\end{eqnarray}
where only two out of three are independent. In other words, one can choose three equations out of (\ref{feom}) $-$ (\ref{eeom3}) as dynamical equations depending on one's convenience. In the present work we choose the three equations (\ref{eeom1}) $-$ (\ref{eeom3}) as the dynamical equations and the remaining one, Eq.\ (\ref{feom}), as the constraint equation.

Next, we eliminate $Y'$ in Eqs.\ (\ref{eeom1}) and (\ref{eeom3}) using differentiation of Eq.\ (\ref{eeom2}) with respect to $r$ and rewrite the two equations after solving the simultaneous equations. Then the equations of motion for $\Phi''$ and $X''$ are obtained as follows:
\begin{equation}
\Phi''=\frac{W_1}{\kappa W}~~~~{\rm and}~~~~X''=\frac{W_2}{W} \,, \label{twoeq}
\end{equation}
where $W_1$, $W_2$, and $W$ are functions of only $X'$\,, $Y$\,, $\Phi$\,, and $\Phi'$ whose detailed expressions are shown in Appendix A.

We first examine the existence of a black hole solution with an event horizon. The event horizon is simply the hypersurface at which $g^{rr}(r_h)=0$ or $g_{rr}(r_h)=\infty$. We check the divergence of the metric function $e^{Y(r)}$ at the event horizon $r_h$. We rearrange the terms in Eq.\ (\ref{eeom2}) to get
\begin{equation}
e^Y = \frac{1}{2} \left[A\pm \sqrt{A^2 +B} \right]\,, \label{eybl}
\end{equation}
where $A=(r-4\alpha\gamma\kappa e^{-\gamma\Phi}\Phi')X' - \frac{1}{2}\kappa r^2 \Phi'^2 + 1 $ and $B= 48\alpha\gamma\kappa e^{-\gamma\Phi}\Phi' X'$. We take the plus sign in Eq.\ (\ref{eybl}).

Assuming $\Phi_h$ and $\Phi'_h$ to be finite makes $X' \rightarrow \infty$ at the horizon, as can be seen from Eq.\ (\ref{eeom2}). We expand the right-hand side of Eq.\ (\ref{eybl}) near the event horizon as follows:
\begin{eqnarray}
e^Y &=&(r-4\alpha\gamma\kappa e^{-\gamma\Phi}\Phi')X' + \frac{[4r+32\alpha\gamma\kappa e^{-\gamma\Phi}\Phi' -2r^3\kappa\Phi'^2 + 8\alpha\gamma\kappa^2r^2e^{-\gamma\Phi}\Phi'^3 ]}{4(r-4\alpha\gamma\kappa e^{-\gamma\Phi}\Phi')} \nonumber \\
&+& {\cal O} \left(\frac{1}{X'} \right)\,, \label{exnearhor}
\end{eqnarray}
where the quantity $(r-4\alpha\gamma\kappa e^{-\gamma\Phi}\Phi')$ is finite.

After substituting Eq.\ (\ref{exnearhor}) in Eq.\ (\ref{twoeq}), we obtain the following:
\begin{eqnarray}
&&\Phi'' = \frac{DH}{\kappa E}X'+ {\cal O} (1) , \label{xd1d00}\\
&&X'' = - \frac{K}{E}X'^2 + {\cal O} (X') + {\cal O} (1)\,,   \label{d2d00}
\end{eqnarray}
where
\begin{eqnarray}
D&=& r-4 \alpha  \gamma  \kappa  e^{-\gamma  \Phi} \Phi' \,, \nonumber \\
H&=& 4\alpha \gamma\kappa^2 r^2 e^{-\gamma \Phi} \Phi'^2 -\kappa r^3 \Phi' + 12 \alpha\gamma\kappa e^{-\gamma\Phi}\,, \nonumber \\
E&=& r^4 - 4 \alpha  \gamma  \kappa  r^3 e^{-\gamma\Phi}\Phi' - 96 \alpha ^2 \gamma^2 \kappa e^{-2 \gamma \Phi }\,, \nonumber \\
K&=& r^4 + 16 \alpha ^2 \gamma ^2 \kappa ^2 r^2 e^{-2 \gamma
   \Phi }\Phi'^2 - 8 \alpha  \gamma  \kappa  r^3 e^{-\gamma\Phi}\Phi' - 48 \alpha ^2 \gamma
   ^2 \kappa  e^{-2 \gamma  \Phi }\,. \nonumber \\
\end{eqnarray}

We check the behaviors of the metric functions and the scalar field at the event horizon $r_h$\,. To keep $\Phi''_h$ finite, we choose $H=0$. Then we can estimate $\Phi''={\cal O} (1)$ from Eq.\ (\ref{xd1d00}) and $X'= \frac{1}{r-r_h} +{\cal O} (1)$ from Eq.\ (\ref{d2d00}). Under $H=0$\,, $\Phi'_h$ is related to $\Phi_h$ as follows:
\begin{equation}
\Phi'_h = \frac{ r_h e^{\gamma\Phi_h}}{8\alpha\gamma\kappa} \left(1 \pm \sqrt{1-192 e^{-2\gamma\Phi_h}\alpha^2\gamma^2\kappa/r^4_h}  \right)\,. \label{phip}
\end{equation}

From the condition that $\Phi'_h$ have real values we obtain the following condition:
\begin{equation}
e^{-\gamma\Phi_h} < \frac{r^2_h}{\alpha}\frac{1}{\gamma\sqrt{192\kappa}}\,. \label{bcphi}
\end{equation}
This is the condition for the existence of a black hole solution with appropriate boundary values $r_h$ and $\Phi_h$ in given parameter values.

The solutions take the asymptotic form
\begin{eqnarray}
e^{X} &\simeq& 1- \frac{2M}{r} + {\cal O} (1/r^3) \,, \label{asymsoltt} \\
\Phi &\simeq& \Phi_{\infty} + \frac{Q}{r} + {\cal O} (1/r^2)\,, \label{asymsolsf}
\end{eqnarray}
where $M$ denotes the ADM mass, $Q$ the scalar charge, and $\Phi_{\infty}$ the asymptotic value of the scalar field, which will be used to rescale the scalar field and radial coordinate in this work. The mass of a hairy black hole is represented as follows \cite{sugo}:
\begin{equation}
M(r)= {M}(r_h) + M_{\rm hair}\,. \label{masshbh}
\end{equation}
where ${M}(r_h)=\frac{1}{2}r_h$ is the mass of a black hole subtracting the contribution coming from the existence of a scalar hair. The second term in the right-hand side, $M_{\rm hair}$, represents the contribution from the scalar hair with $\rho^{GB}$ coming from the DEGB term. The $M(r)$ increases up to some constant as the distance from the horizon increases, if $\Phi'$ and $\rho^{GB}$ rapidly decrease to zero.

\subsection{Numerical Construction of Black Holes }

We obtain the DEGB black hole by solving Eqs.~(\ref{eeom2}) and (\ref{twoeq})\,, generally.
\begin{figure}[h]
\begin{center}
\subfigure []{\includegraphics[width=3.04in]{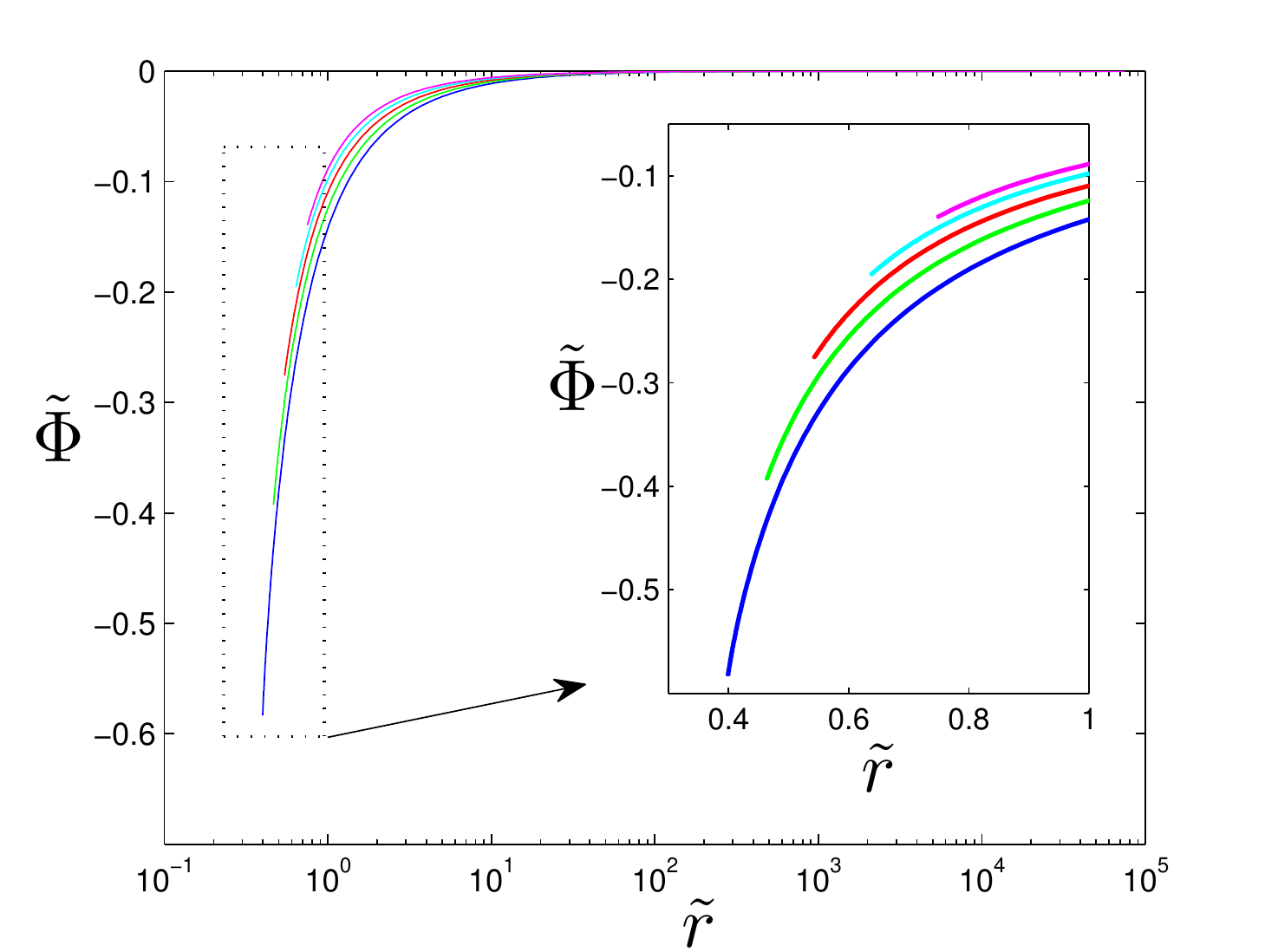}}
\subfigure []{\includegraphics[width=3.3in]{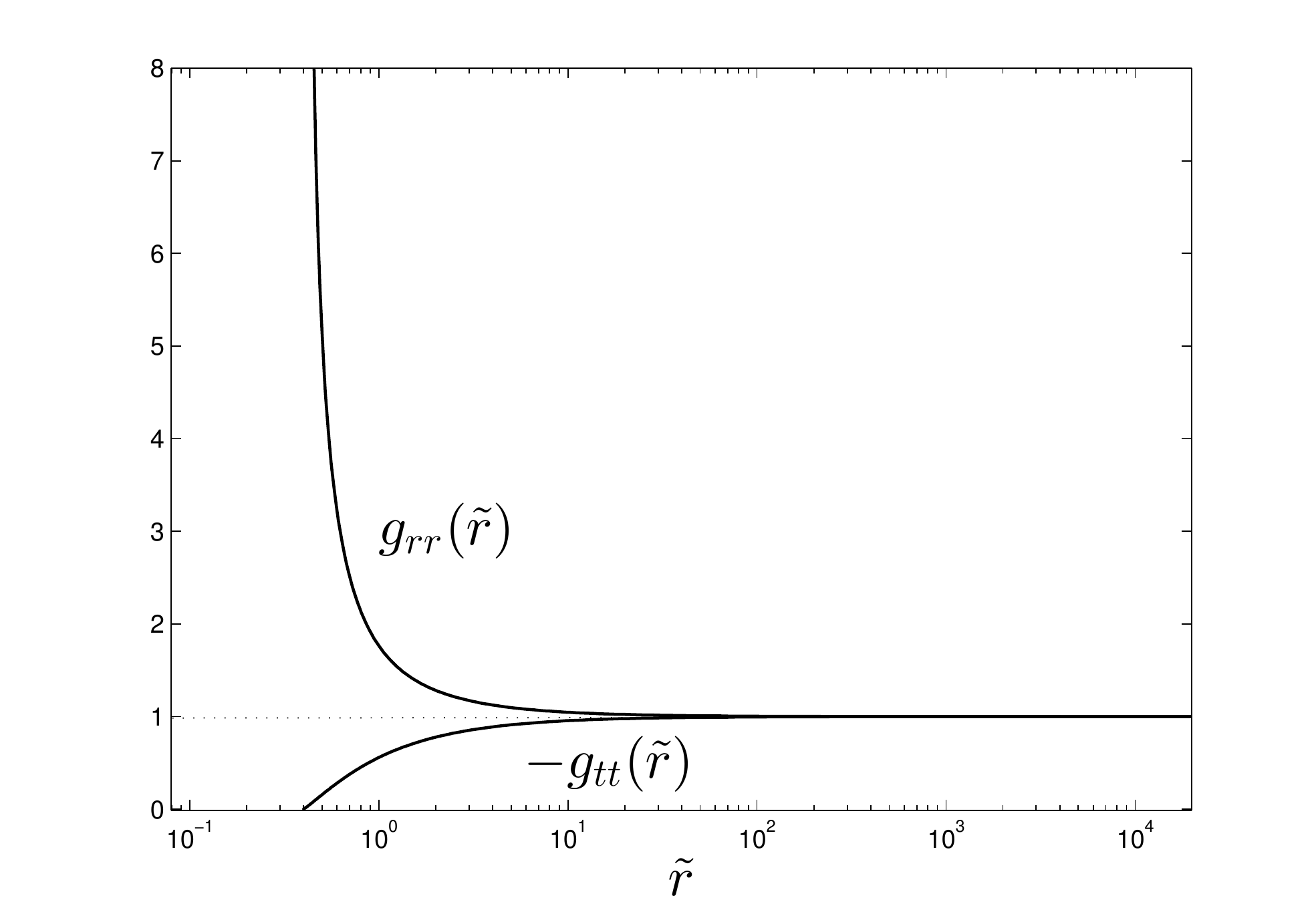}}
\vspace{-0.3cm}
\caption{\footnotesize{(a) Scalar field profiles for radial coordinate in $\gamma=1/6$,  and $\alpha=1/16$. The five solid lines correspond to different DEGB black hole solutions. (b) the numerical solutions represent the metric components $g_{tt}$ and $g_{rr}$ for $r_{h}=1$.}}\label{fig:hairymetric}
\end{center}
\end{figure}
We impose the initial conditions as follows: We first fix the couplings $\alpha$ and $\gamma$ in DEGB theory. For a hairy black hole having $r_h$\,, the maximum value of $\Phi_h$ saturates the inequality Eq.\ (\ref{bcphi}). Hairy black hole solutions exist for $\Phi_h$ less than the maximum $\Phi_h$ value. $\Phi'_h$ is obtained from Eq.\ (\ref{phip}). The initial value of $X'$ is obtained from the relation $X'= \frac{1}{r-r_h}$. We choose initial values $r=r_h+\epsilon$ where ${\epsilon}=10^{-10}$\,. The initial value of $Y$ is obtained from Eq.\ (\ref{eeom2}). The initial value of $X$ is obtained from the equality $Y=-X$ to be satisfied in the asymptotic region. The equations are integrated through the $4th$-order Runge-Kutta-Fehlberg method from $r_{h}$ to $r \rightarrow \infty$\,. Our calculations are for the relative tolerance of $10^{-8}$ and the absolute tolerance of $10^{-8}$\,. The ADM mass $2M$ is obtained from Eq.~(\ref{asymsoltt})\,.

\begin{figure}[h]
\begin{center}
\subfigure []{\includegraphics[width=3.0in]{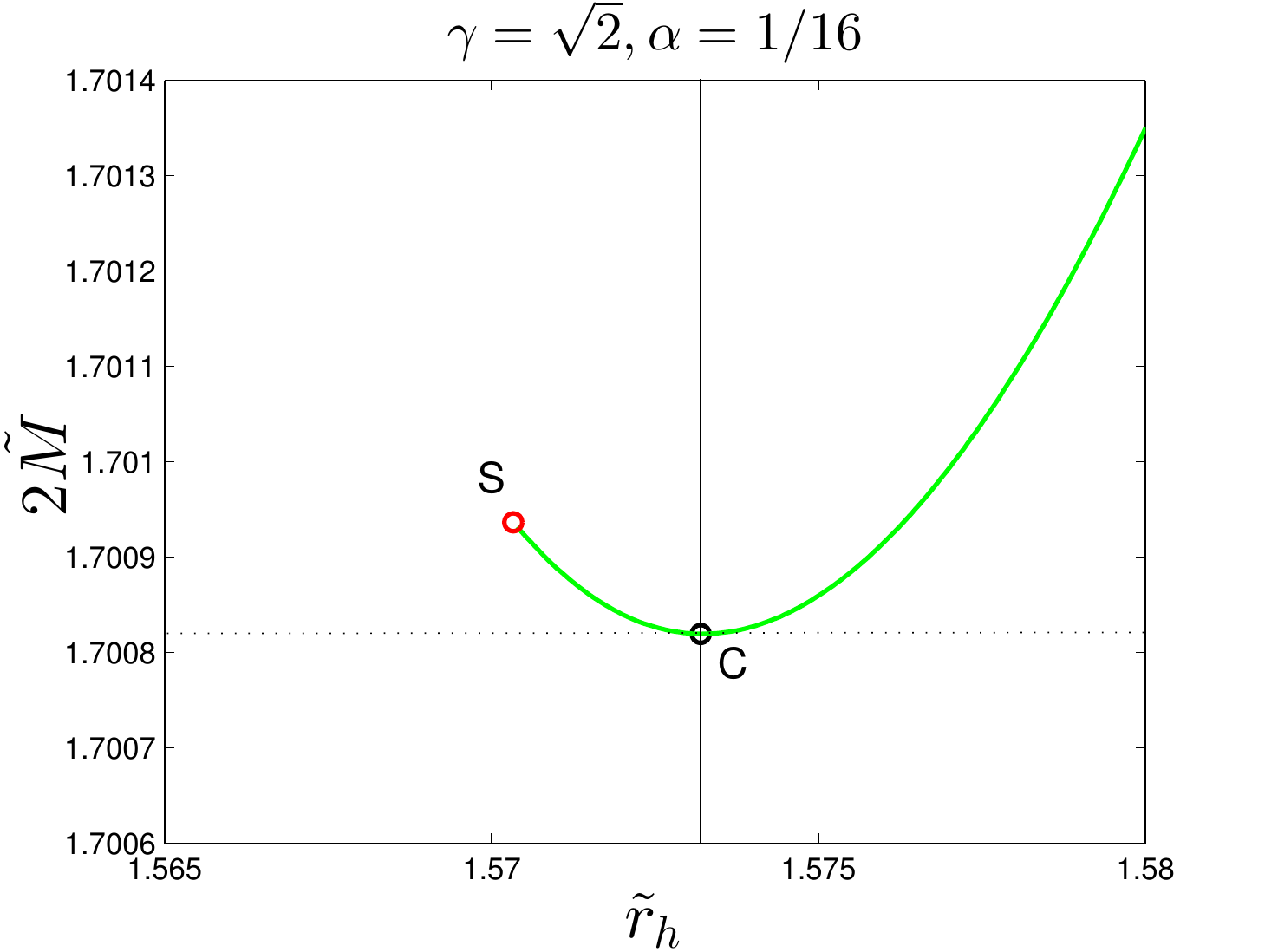}}
\subfigure []{\includegraphics[width=3.0in]{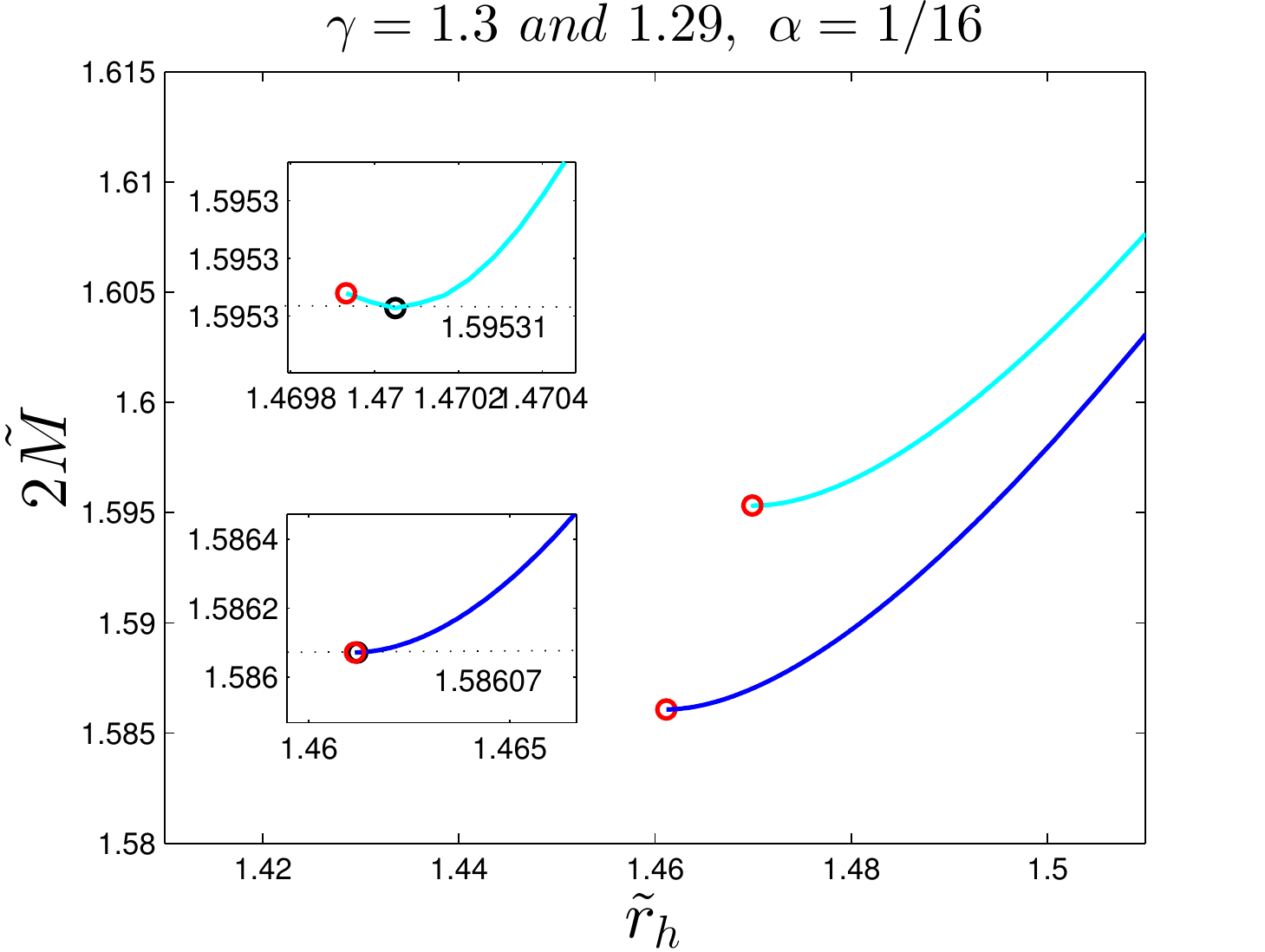}}
\vspace{-0.3cm}
\caption{\footnotesize{Coupling $\gamma$ dependency of the minimum mass for fixed $\alpha=\frac{1}{16}$. For the coupling $\alpha=1/16$, (a) Singular point $S$ and the minimum mass $C$ exist for $\gamma=\sqrt{2}$(green). (b) The singular point $S$ coincides with the point $C$ between $\gamma=1.29$(blue) and $1.30$(cyan). There is no lower branch below $\gamma=1.29$.}}\label{fig:branch2}
\end{center}
\end{figure}
The scalar field $\Phi$ should be asymptotically flat, so the values of the scalar fields can be set to $\tilde{\Phi}_{\infty}=0$ in the asymptotic region. Under this condition, we redefine $\Phi$ by ${\tilde{\Phi}}=\Phi-\Phi_{\infty}$\,. To make the equations of motion invariant under this field shift, the radial coordinate is rescaled to $r \rightarrow {\tilde{r}}=r e^{\gamma \Phi_{\infty}/2}$\,. In the rescaled system, the mass $M$ and charge $Q$ are also rewritten $M \rightarrow {\tilde{M}}=M e^{\gamma \Phi_{\infty}/2} $ and $Q \rightarrow {\tilde{Q}}=Q e^{\gamma \Phi_{\infty}/2}$, respectively. The other parameters are not changed on the rescaled system. Under this rescaling, the whole solution curve $(r_h, \Phi_h)$ corresponds to the unique solution line $(\tilde{r}_h, \tilde{\Phi}_h)$. Therefore, even if we obtain black hole solutions for the specific horizon $r_h$ with couplings $\alpha$ and $\gamma$, the solution becomes the same solution as in the rescaled system. Every choice on $r_h$ leads to the same solution, so our solution is universal. The detailed discussions are in Appendix B. This rescaled system still satisfies the equations of motions in Eqs.~(\ref{feom})\,, (\ref{eeom1}), (\ref{eeom2}), and (\ref{eeom3}) as well as the boundary condition in Eq.\ (\ref{bcphi})\,. We choose the parameter $\kappa=1$ for convenience without loss of generality. From now on, we use rescaled variables.

\begin{figure}[h]
\begin{center}
\includegraphics[width=3.5in]{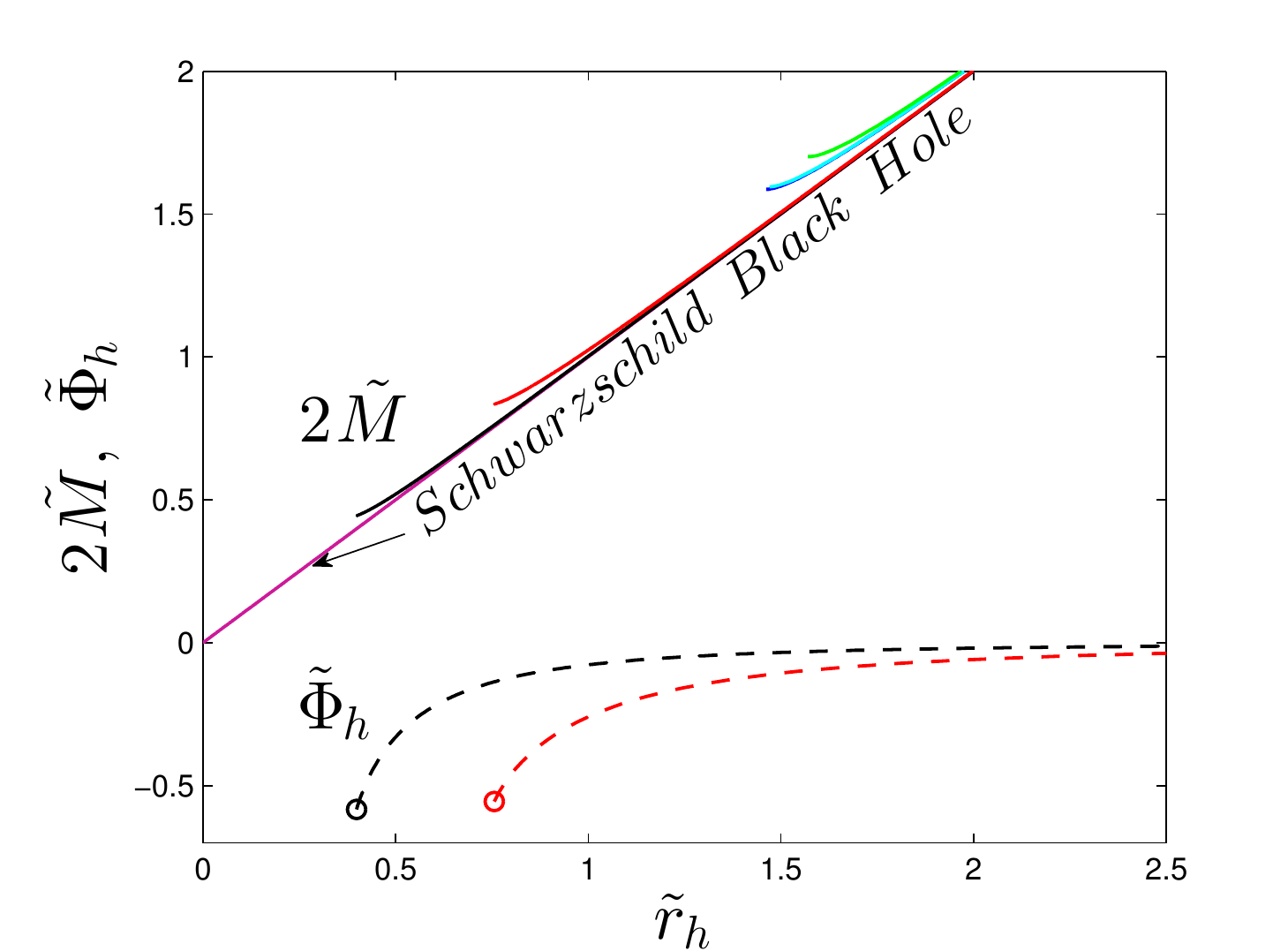}
\end{center}
\vspace{-0.5cm}
\caption{\footnotesize{The black hole mass $\tilde{M}$(solid) and $\tilde{\Phi}_h$(dashed) with respect to $\tilde{r}_h$ for $\gamma=\sqrt{2}$(green), $\gamma=1.3$(cyan), $\gamma=1.29$(blue), $\gamma=1/2$(red), $\gamma=1/6$(black) and $\gamma=0$(puple). In the limit of coupling $\gamma\rightarrow 0$, the solution approaches the Schwarzschild cases.}} \label{fig:tmr}
\end{figure}
The scalar field $\tilde{\Phi}$ is obtained for given couplings and black hole horizons, as shown in Fig.~\ref{fig:hairymetric}(a)\,. The scalar field profiles start at negative values $\tilde{\Phi}_h$ at the black hole horizon $\tilde{r}_h$ monotonically approaching to zero. The bottom profile of the blue solid line shows the possible minimum horizon radius and maximum magnitude scalar field $|\tilde{\Phi}_h|$ which saturates inequality in Eq.~(\ref{bcphi}) for given couplings $\alpha$ and $\gamma$\,. The upper lines satisfy the inequality in Eq.~(\ref{bcphi}). If DEGB black hole horizon becomes larger, the magnitude of the scalar field becomes smaller. In the large horizon radius limit, the scalar field approaches to zero, and then the black hole becomes a Schwarzschild black hole. Fig.~\ref{fig:hairymetric} shows that the metric component $g_{rr}$ becomes infinite at the horizon, while $g_{tt}$ is approaching zero, whereas both metrics are asymptotically approaching the value $1$.

For fixed $\alpha$, the singular point $S$ and minimum mass $C$ exist for large $\gamma$ as shown in Fig.~\ref{fig:branch2}(a). There exists a minimum mass $\tilde{M}_{min}$ at the extremal point $C$ as shown in Fig.~\ref{fig:branch2}(a)\cite{Kanti:1995vq,Torii:1996yi,Kanti:1997br,Torii:1998gm}. For small $\gamma$, the singular point $S$ gets closer to the minimum mass point $C$ as shown in Fig.~\ref{fig:branch2}(b). The solutions between point $S$ and $C$ in Fig.~\ref{fig:branch2}(a) are unstable for perturbations and end at the singular point $S$ which saturates to equality in Eq.~(\ref{bcphi}). In other words, there are two black holes for a given mass in which the smaller one is unstable under perturbations. Below $\gamma=1.29$, the solutions are perturbatively stable and approach the Schwarzschild black hole in the limit of $\gamma$ going to zero. These solutions depend on couplings $\gamma$ as shown in Fig.~\ref{fig:branch2}. We will investigate these solutions under fragmentation below $\gamma=1.29$.

In Fig.~\ref{fig:tmr}\,, the black hole mass $\tilde{M}$ and hair $\tilde{\Phi}_h$ are plotted for different values of the coupling $\gamma$ values. Each point describes the black hole mass $\tilde{M}$(solid lines) and $\tilde{\Phi}_h$(dashed lines) for a given horizon $\tilde{r}_h$. The black hole mass monotonically increases with respect to $\tilde{r}_h$\,. The dilaton field magnitude $\tilde{\Phi}_h$ monotonically decreases with respect to $\tilde{r}_h$\,. For given $\gamma$, the black hole mass $\tilde{M}$ and $\tilde{\Phi}_h$ have minimum values saturating inequality Eq.~(\ref{bcphi}). When the coupling $\gamma$ decreases in Fig.~\ref{fig:tmr}, both lines move down at the left-hand side. Eventually, the solutions become those of EGB theory in the limit of $\gamma \rightarrow 0$, so the mass profile with respect to $\tilde{r}_h$ should approach the line of $\tilde{r}_h=0$. The graph $\tilde{M}$ is proportional to $\sqrt{\alpha}$ as mentioned.

\section{Instability from fragmentation \label{sec3}}

Black holes may undergo instability at some couplings and break apart into black holes~\cite{Emparan:2003wm}\,. The initial phase is a single black hole having mass $\tilde{M}$\,, which is the function of an initial horizon $\tilde{r}_{h}$. The final phase is two black holes far from each other. One of these black holes has a mass $\tilde{m}$ and a linear momentum $P_1$, and the other has $\tilde{M}-\tilde{m}$ and $P_{2}$ under mass and momentum conservation. The total linear momentum is zero in the initial and final phases. This final phase is specified by a mass ratio $\delta=\frac{\tilde{m}}{\tilde{M}}$. We denote the final phase as $(\delta,1-\delta)$. The maximum value of $\delta$ is $\frac{1}{2}$ for half fragmentation. The possible minimum mass ratio $\bar{\delta}$ is given as $\frac{\tilde{M}_{min}}{\tilde{M}}$. The minimum mass ratio $\bar{\delta}$ has a finite value for a DEGB black hole, because the black hole has minimum mass $\tilde{M}_{min}$. The black holes can be fragmented only when it exceeds twice the minimum mass. With a black hole mass below twice the minimum mass, there are no fragmented black hole solutions, so these black holes are absolutely stable. The mass and momenta of the black hole are related\,\cite{Gwak:2010in},
\begin{eqnarray}
\tilde{M}=\sqrt{(\delta \tilde{M})^2+P_1^2}+\sqrt{(1-\delta)^2 \tilde{M}^2+P_{2}^2}\,.
\end{eqnarray}
The linear momenta are arbitrary, so we set $P_1=P_2=0$ to maximize the total entropy of the final phase. In this condition, the black hole slightly breaks into two black holes with negligible momenta. The initial phase decays to the final phase if the final entropy is larger than that of the initial phase.

The entropy of the initial phase $S_i$ is that of one DEGB black hole. The black hole entropy with the form of a polynomial of the Ricci scalar is given as
\begin{eqnarray}
\label{wald}
S=-2\pi \int_{\Sigma} E^{\mu\nu\rho\sigma}_R \epsilon_{\mu\nu}\epsilon_{\rho\sigma}\,,\,\,E^{\mu\nu\rho\sigma}_R=\frac{\partial\mathcal{L}}{\partial R_{\mu\nu\rho\sigma}}\,,
\end{eqnarray}
where $\Sigma$\,, $\epsilon_{\mu\nu}\,$, and $\mathcal{L}$\,, are the bifurcation horizon 2-surface,the  volume element binormal, and the Lagrangian density~\cite{Jacobson:1995uq,Wald:1993nt,Chatterjee:2013daa}\,. The initial DEGB black hole entropy~\cite{Torii:1998gm} is
\begin{eqnarray}
\label{DEGBentropy}
S_i=\frac{\pi\tilde{r}_h^2}{G} \left(1+\frac{8\alpha\kappa}{\tilde{r}_h^2}e^{-\gamma\tilde{\Phi}_{h}}\right)\,,
\end{eqnarray}
where the DEGB black hole entropy has an additional term from the hair contribution. This additional term exists in the Euclidean path integral \cite{giha}. The entropy can be obtained from the relation $S= \beta E - I_E$, in which $\beta$ is the inverse of the temperature, $E$ is the energy or the mass, and $I_E$ is the Euclidean action. For the Schwarzschild black hole in Einstein gravity, the quantity $I_E$ has only a contribution coming from the boundary term, $\frac{\beta E}{2}=\frac{A}{4G}$. For the black hole in EGB and DEGB theory, there is a non-vanishing contribution from the bulk term with the higher-curvature GB term in the Euclidean action. The non-vanishing contribution gives rise to the additional entropy correction Eq.\ (\ref{DEGBentropy}).

After fragmentation, we expect that two black holes are far from each other in the final phase. Therefore, we suppose that the black holes do not interact as if they were independent spacetimes. In this case, the black hole entropy in the final phases is approximately described by the simple sum of two fragmentated black holes. Precisely, if we treat two fragmentation black holes with interaction, the final phase entropy should include also an interaction term instead of only a simple sum. In the Euclidean path integral, each entropy of a black hole has the contributions coming from both the bulk and the boundary term. Thus we have added both contributions for the two black holes, which will cause the fragmentation instability.

First, as the simplest case, we study the possible fragmentation of a Schwarzschild black hole. The fragmentation instability depends on the mass ratio $\delta$ between the initial and the final phase of the black hole mass. For the case of $(\delta, 1-\delta)$, the entropy ratio is given as
\begin{eqnarray}
\frac{S_f}{S_i}=\frac{(\delta\,\tilde{r}_h)^2+((1-\delta)\,\tilde{r}_h)^2}{\tilde{r}_h^2}=\delta^2+(1-\delta)^2\,,
\end{eqnarray}
where we denote the initial and final phase entropy $S_i$ and $S_f$. The entropy ratio is always smaller than $1$, so the entropy of a initial phase is larger than that of a final phase. Therefore, a Schwarzschild black hole is always stable under fragmentation. The entropy ratio marginally approaches $1$ in $\delta \rightarrow 0$, because the entropy is proportional to the square of the horizon radius, while the mass is proportional to the horizon radius.

\begin{figure}[h]
\begin{center}
\subfigure []{\includegraphics[width=3in]{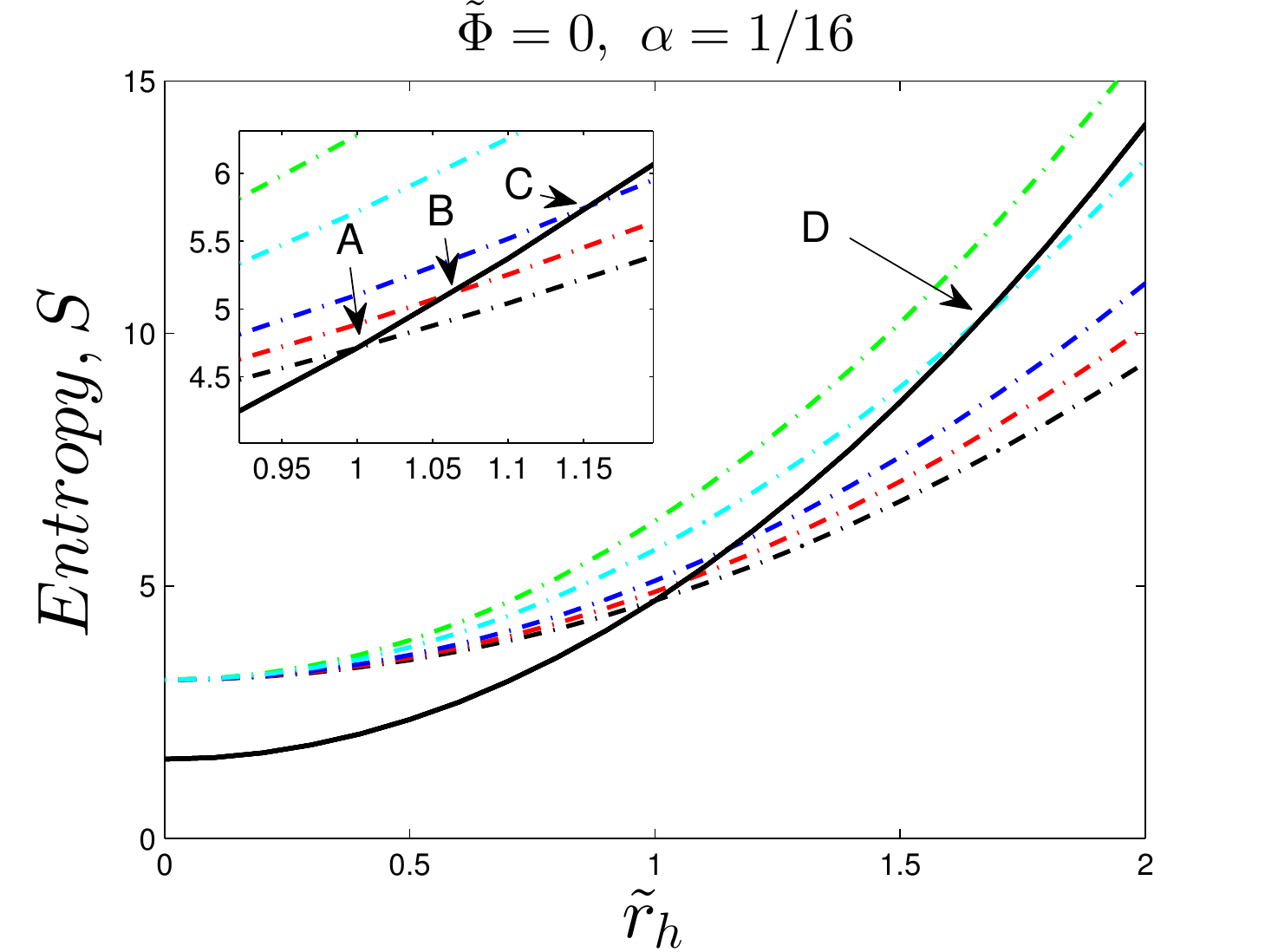}}
\subfigure []{\includegraphics[width=3in]{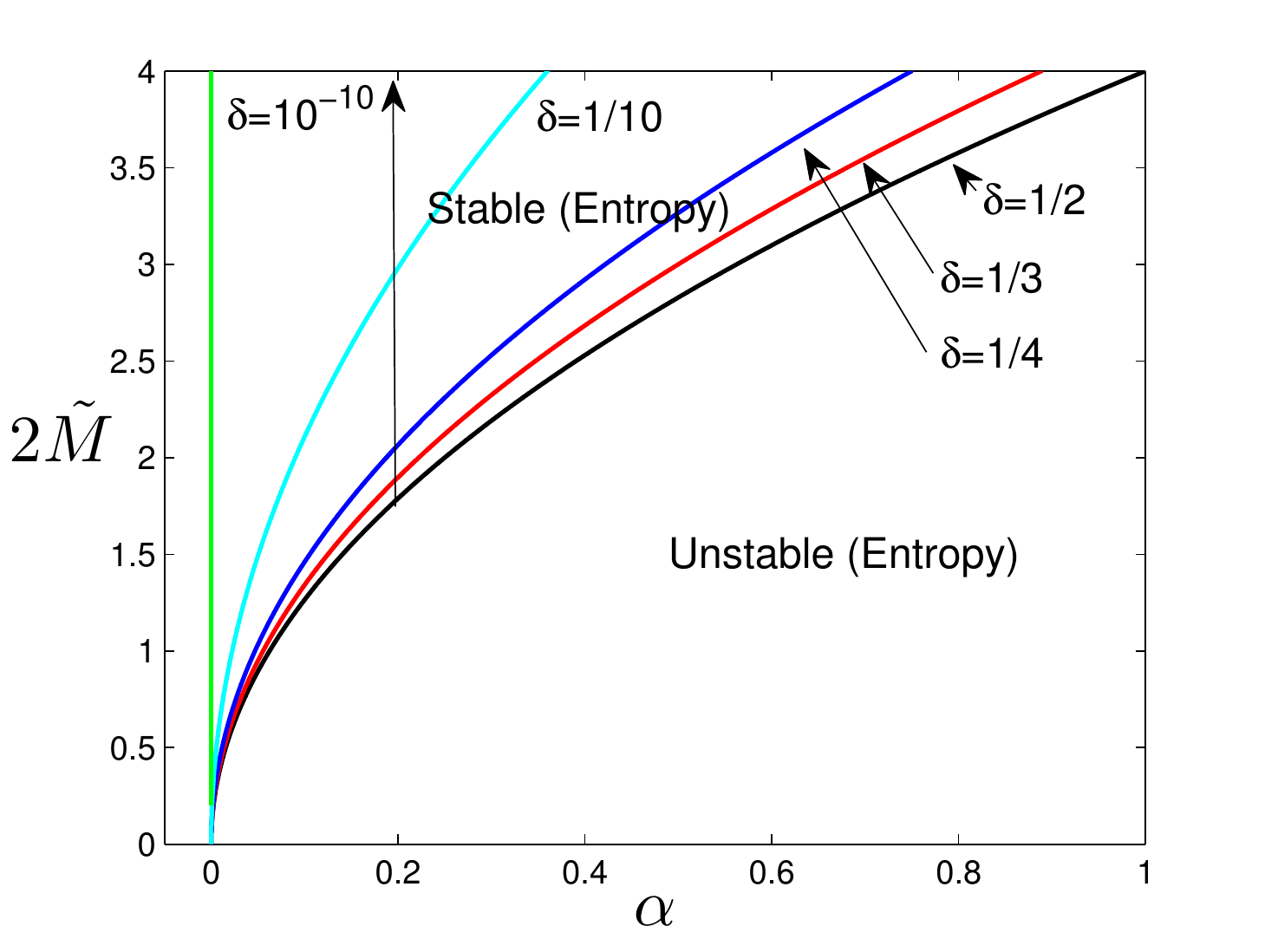}}
\end{center}
\vspace{-0.6cm}
\caption{\footnotesize{(a) Fragmentation ratio and EGB black hole entropy. The black solid line is the initial phase entropy. The black, red, blue, cyan, and green dashed-dot lines are the cases of $(\frac{1}{2},\frac{1}{2})$, $(\frac{1}{3},\frac{2}{3})$, $(\frac{1}{4},\frac{3}{4})$, $(\frac{1}{10},\frac{9}{10})$, and $(10^{-10},1-10^{-10})$. The crossing points go up from point $A$ to $D$ with changing $\delta$. We fix $\kappa=1$. (b) Phase diagram of EGB black hole for ${\delta}=1/2$ (black solid line), ${\delta}=1/3$ (red solid line), ${\delta}=1/4$ (blue solid line), ${\delta}=1/10$ (cyan solid line) and ${\delta}=10^{-10}$ (green solid line) fragments corresponding to crossing points between the initial and final phase of the black hole entropy; each color of the lines is for the same as those in figure (a).
}}\label{fig:estimatefrag}
\end{figure}

These phenomena become different in the theory with the higher order of curvature term. In the EGB theory of $\gamma=0$, the static black hole metric is the same as that of a Schwarzschild solution and exists for arbitrary mass. However, the entropy has a quantum correction coming from GB term of the higher order of curvature. The initial black hole entropy is
\begin{eqnarray}
S_i=\frac{A_H}{4G}\left(1+\frac{8\alpha\kappa}{\tilde{r}_h^2}\right)=\frac{\pi}{G}\left(\tilde{r}_h^2+8\alpha\kappa\right)\,.
\label{eq:DEGBent1}
\end{eqnarray}
Unlike Schwarzschild black holes, the fragmentation instability occurs depending on the fragmentation ratio $\delta$. For the case of $(\delta, 1-\delta)$ fragmentation, the final phase entropy is given
\begin{eqnarray}
S_f=\frac{\pi}{G}\left((\delta \tilde{r}_h)^2+8\alpha\kappa\right)+\frac{\pi}{G}\left(((1-\delta) \tilde{r}_h)^2+8\alpha\kappa\right)\,,
\end{eqnarray}
where the initial and final phases are connected under the quantum or thermal fluctuation, which can allow the topology changing process. Exactly, the entropy contribution of the GB term in the final phase is not twice of that in initial phase. Also, that is not same one for a black hole in the initial phase. This is because the action integral of the GB term is an invariant quantity which provides the information on the topology of that spacetime manifold. The symmetry of the initial black hole spacetime is changed or broken into that of two black holes. Furthermore, black holes include the binding energy between them. In order to obtain the exact prescription, we should solve the difficult non-linear equations of motions in the DEGB theory. However, our goal is to investigate this phenomena thermodynamically. To simplify this problem, we assume that the fragmented black holes are far each other, thus the black holes could exist independently in the spacetime with each asymptotic boundary and we could ignore the binding energy between black holes. The EGB black hole is unstable if,
\begin{eqnarray}
\frac{S_f}{S_i}=\frac{\left((\delta \tilde{r}_h)^2+8\alpha\kappa\right)+\left(((1-\delta) \tilde{r}_h)^2+8\alpha\kappa\right)}{\left(\tilde{r}_h^2+8\alpha\kappa\right)}>1\,.
\end{eqnarray}
In a small mass limit, the ratio becomes $2$ from the dominant correction term. On the other hand, in a large mass limit, the entropy ratio becomes $\delta^2+(1-\delta)^2$ as same as that of a Schwarzschild case. Therefore, the EGB black hole with a small mass is unstable, while a massive EGB black hole is stable. There exists a crossing point between initial and final phase entropy. The crossing points are obtained from $S_f/S_i=1$,
\begin{eqnarray}
\tilde{r}_{cross}=2\sqrt{\frac{\alpha\kappa}{\delta(1-\delta)}}\label{deltaent1}\,.
\end{eqnarray}
For given parameter, EGB black holes are unstable below $\tilde{r}_{cross}$. There is no minimum mass of the EGB black hole, so mass ratio $\delta$ has a range of $0<\delta<\frac{1}{2}$. Several initial and final phase entropies for mass ratios are shown in Fig.~\ref{fig:estimatefrag}(a). The smaller mass ratio covers larger mass range as shown in Fig.~\ref{fig:estimatefrag}(a). Overall behaviors of entropy are independent on mass ratio as same as MP black hole cases\cite{Emparan:2003wm,Gwak:2014xra}\,.
Each mass ratio $\delta$ leads to each line in the phase diagram for given $\alpha$ as shown in Fig.\ \ref{fig:estimatefrag}(b). The mass ratio can have continuous values, and the black hole has stable and unstable phases. The minimum unstable region is at $\delta=\frac{1}{2}$. For the limit of $\delta\rightarrow 0$, all of the EGB black holes become unstable for fragmentations, as shown in Fig.\ \ref{fig:estimatefrag}(b).

The DEGB black hole entropy ratio between the initial and the final entropy including the higher-curvature corrections is in the approximation $\tilde{r}_h\approx 2\tilde{M}$,
\begin{eqnarray}
\label{entropyratio}
\frac{S_f}{S_i}=\frac{\left((\delta \tilde{r}_{h})^2+8\alpha\kappa e^{-\gamma\tilde{\Phi}_{\delta}}\right)+\left(((1-\delta)\tilde{r}_{h})^2+8\alpha\kappa e^{-\gamma\tilde{\Phi}_{1-\delta}}\right)}{\left(\tilde{r}_{h}^2+8\alpha\kappa e^{-\gamma\tilde{\Phi}_{h}}\right)}\,,
\end{eqnarray}
where $\tilde{\Phi}_{h}$, $\tilde{\Phi}_{\delta}$, and $\tilde{\Phi}_{1-\delta}$ are the scalar field values at the initial and final black hole horizon. In the large mass limit $\tilde{r}_h\gg 1$, the entropy ratio becomes that of the Schwarschild case,
\begin{eqnarray}
\frac{S_f}{S_i}=\delta^2+(1-\delta)^2<1.
\end{eqnarray}
Thus, massive DEGB black holes are stable under fragmentation. The small mass limits are bounded to $\tilde{M}_{min}$. DEGB black holes of mass $\tilde{M}_{min}$ are absolutely stable, because there are no fragmented black hole solutions. For values larger than $\tilde{M}_{min}$, the black hole stability is dependent on an entropy correction term. The entropy ratio is given
\begin{eqnarray}
\label{entropyratio2}
\frac{S_f}{S_i}=\frac{{\delta}^2+({\delta}-1)^2+\frac{8\alpha\kappa e^{-\gamma\tilde{\Phi}_{{\delta}}}+8\alpha\kappa e^{-\gamma\tilde{\Phi}_{{1-\delta}}}}{\tilde{r}_h^2}}{1+\frac{8\alpha\kappa e^{-\gamma\tilde{\Phi}_{h}}}{\tilde{r}_h^2}}\,,
\end{eqnarray}
where the horizon radius square term is important in the small black hole. The entropy ratio may increase for a smaller mass like the EGB black holes, but there is an ambiguity since the DEGB black holes have a minimum mass. In this case, there is no proper approximation to describe the instabilities of small mass DEGB black holes. This should become clear through numerical calculation. Also, the minimum mass bounds the fragmentation mass ratio. It is not seen in the Schwarzschild black hole or EGB black hole. The DEGB black holes have more variety as regards properties and behaviors. We will obtain detailed behaviors through numerical calculations.

\section{Numerical analysis for fragmentation instability \label{sec4}}

\begin{figure}[h]
\begin{center}
\centering
\subfigure []{\includegraphics[width=2.in,keepaspectratio]{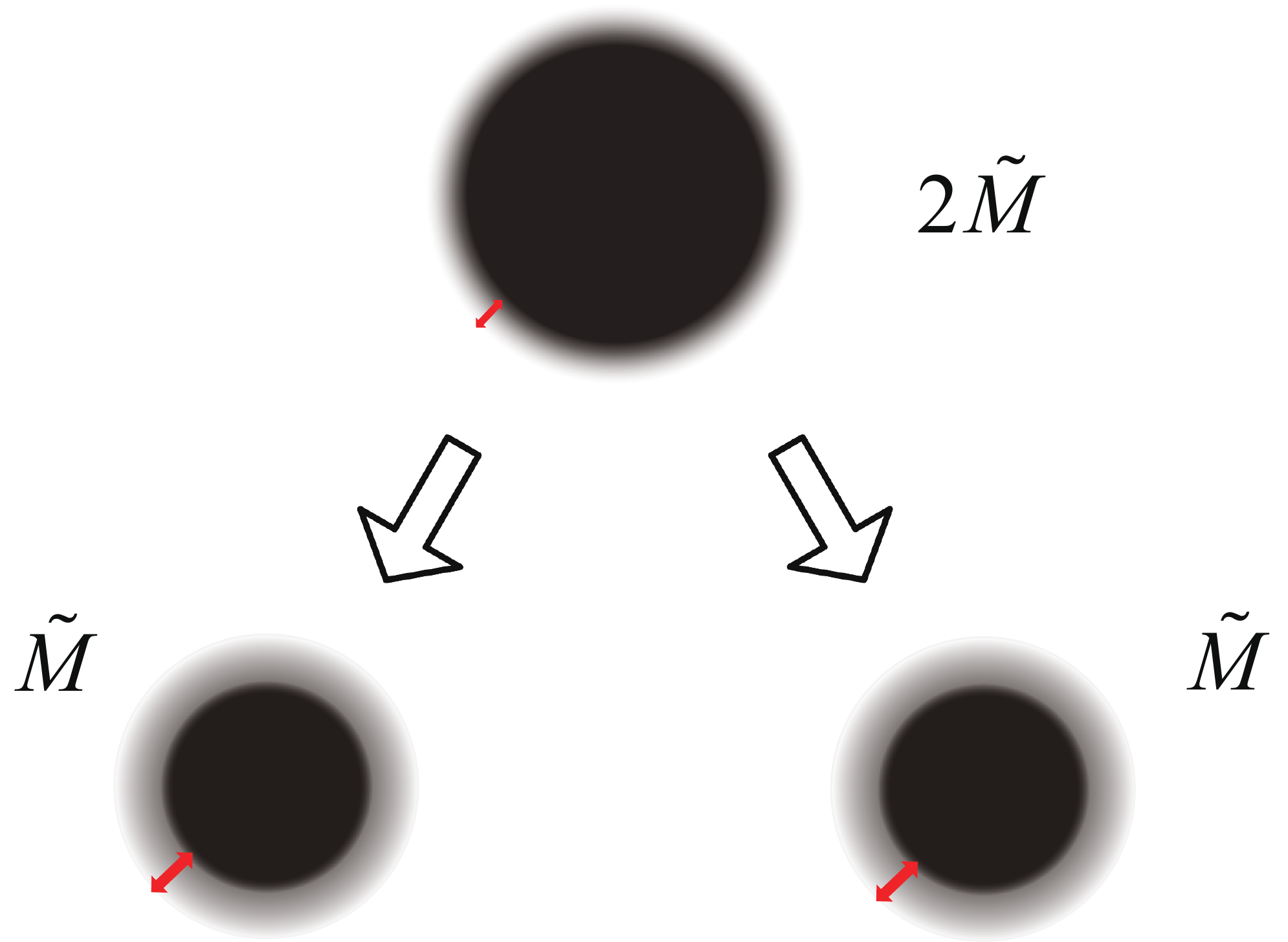}}\hspace{1.2cm}
\subfigure []{\includegraphics[width=2.in,keepaspectratio]{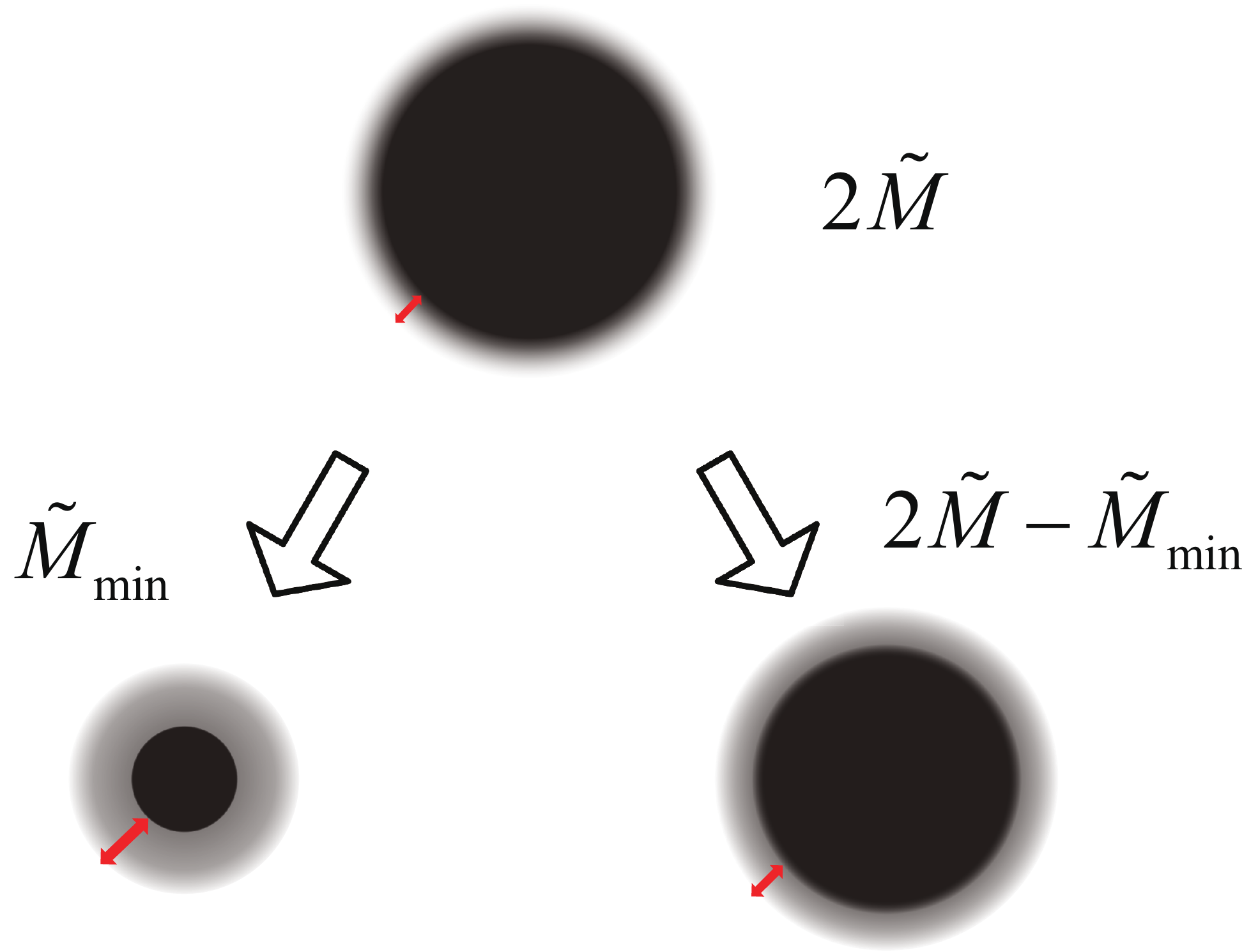}}
\end{center}
\vspace{-0.6cm}
\caption{\footnotesize{Schematic illustration of the fragmentation of hairy black hole from initial one mass $2\tilde{M}$ to final two masses. The blurred gray areas of each black hole present that the hairy profiles exist outside event horizon of the black holes. (a) Fragmentation to $\tilde{M}$ and $\tilde{M}$ (b) Fragmentation to $\tilde{M}_{min}$ and $2\tilde{M}-\tilde{M}_{min}$.}} \label{fig:bhf}
\end{figure}

We investigate the fragmentation instability using a numerical analysis. We consider the fragmentation cases of $\bar{\delta}\leq\delta\leq\frac{1}{2}$ as shown in Fig.~\ref{fig:bhf}.

\begin{figure}[h]
\begin{center}
\subfigure []{\includegraphics[width=2.5in]{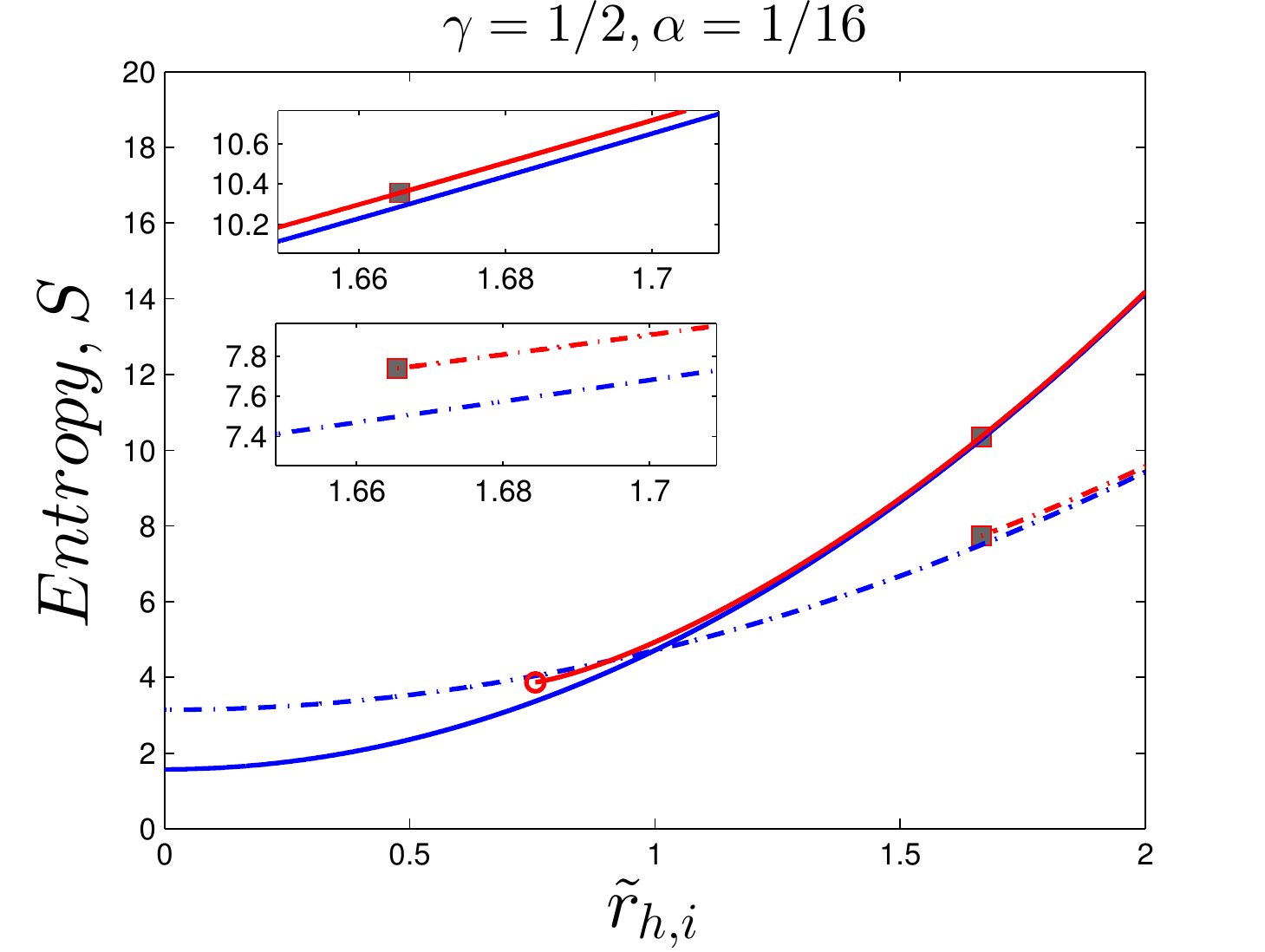}}
\subfigure []{\includegraphics[width=2.5in]{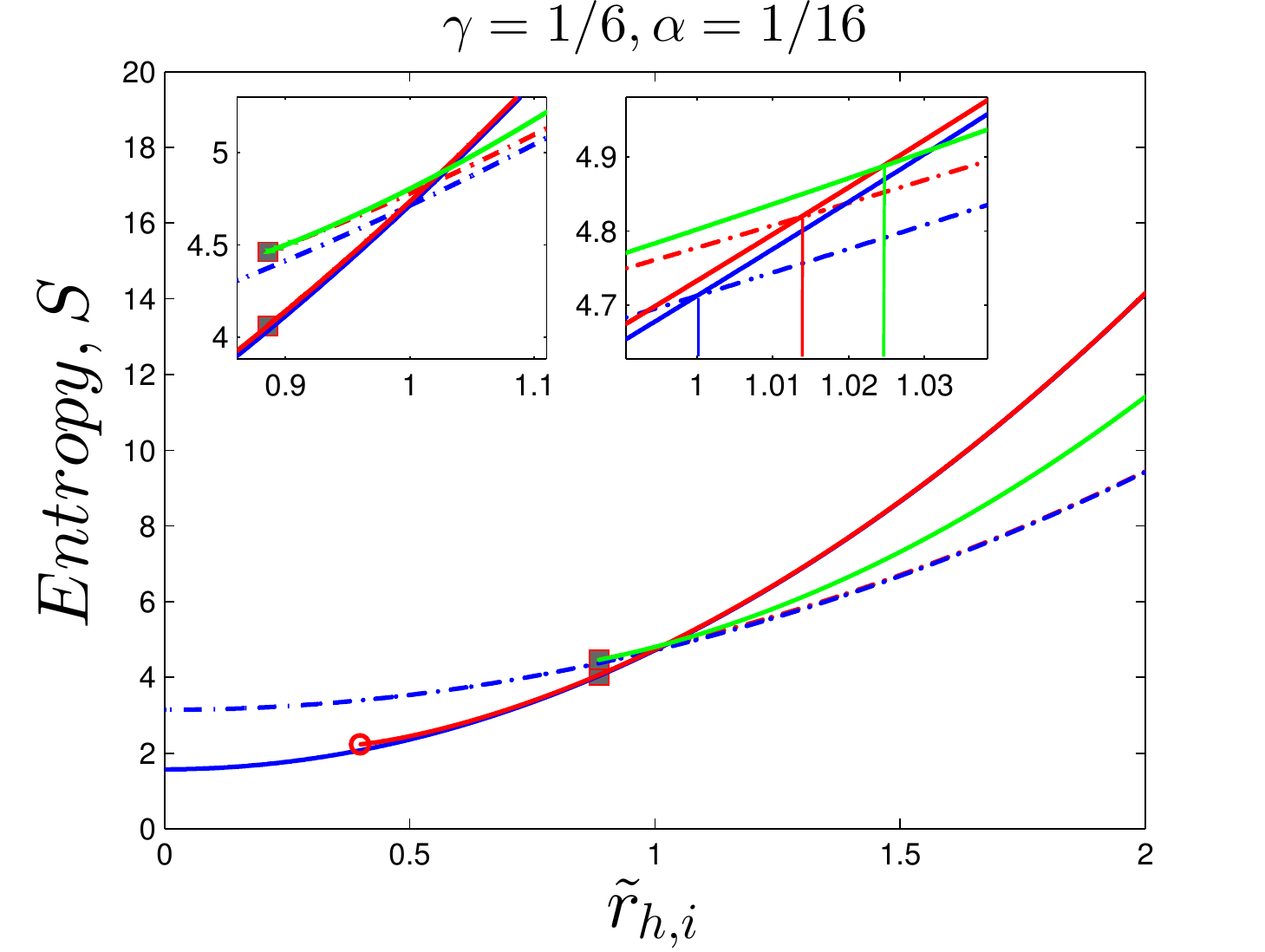}}
\end{center}
\vspace{-0.6cm}
\caption{\footnotesize{The initial and final phase entropies with respect to $r_{h,i}$ for the given couplings $\gamma$ and $\alpha$. The blue solid line and blue dashed-dot line are initial and final phase entropies in EGB theory as a reference for $(\frac{1}{2},\frac{1}{2})$. The red solid line and red dashed-dot line are initial and final phase entropies in DEGB theory for $(\frac{1}{2},\frac{1}{2})$. The initial phase exists above red circle for the minimum mass. The final phase exists above red box for $(\frac{1}{2},\frac{1}{2})$. The green solid line represents fragmentation for marginal mass ratio $\bar{\delta}$.}} \label{fig:estimate}
\end{figure}

The DEGB black hole entropies in Eq.~(\ref{DEGBentropy}) are shown with respect to the horizon radius in Fig.~\ref{fig:estimate} for $\gamma=\frac{1}{2}$ and $\gamma=\frac{1}{6}$ with $\alpha=\frac{1}{16}$. Unlike EGB black holes for a blue line, DEGB black holes have a minimum mass $\tilde{M}_{min}$ for given parameters.
A Red circle corresponds to the initial black hole having a minimum mass. Below the minimum mass, there is no DEGB black hole solution region. The red box corresponds to a fragmented black hole having a minimum mass. The overall behaviors are similar to those of EGB black holes. The DEGB black hole entropy is slightly larger than that of EGB theory, because of the hair contribution. Possible fragmentation of DEGB black holes occurs at twice the minimum mass with $(\frac{1}{2},\frac{1}{2})$ mass ratio. Below half fragmentation, massive DEGB black holes are in the stable(mass) region between red circle and box. In this range, these black holes have no final phase solutions corresponding to decay. In Fig.~\ref{fig:estimate}(a), the initial phases are in the stable(entropy) region, because all final phase entropies are smaller than that of the initial phase above red box. However, in Fig.~\ref{fig:estimate}(b), a red box is located under the crossing point, so DEGB black holes are in the unstable(entropy) region between the red box and the crossing point. Also, above the crossing point, the initial phases are in the stable(entropy) region. For the limit of $\gamma\rightarrow 0$, the red box must approach $\tilde{r}_h=0$. DEGB black holes are more unstable for the smaller mass ratio $\delta$ as for the EGB black hole cases. The largest unstable(entropy) region is given at $\bar{\delta}$. This fragmentation always starts at the $(\frac{1}{2},\frac{1}{2})$ mass ratio and then appears above it, as shown in Fig.~\ref{fig:estimate}(b). The crossing point from $(\bar{\delta},1-\bar{\delta})$ fragmentation appears for a larger initial black hole mass than that of the EGB black hole. As a result, DEGB black holes are stable in a larger mass range.

\begin{figure}[h]
\begin{center}
\subfigure []{\includegraphics[width=2.5in]{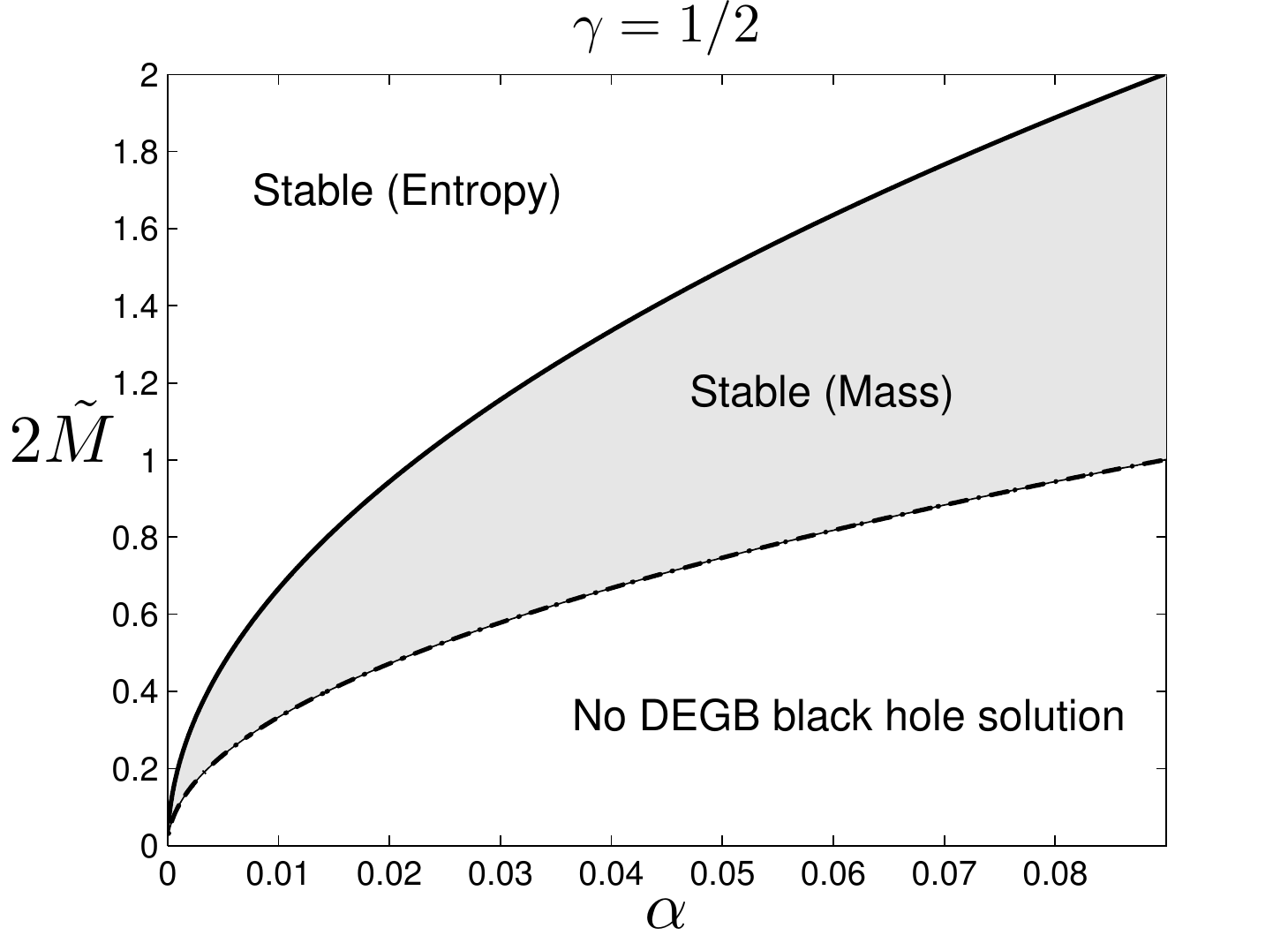}}
\subfigure []{\includegraphics[width=2.5in]{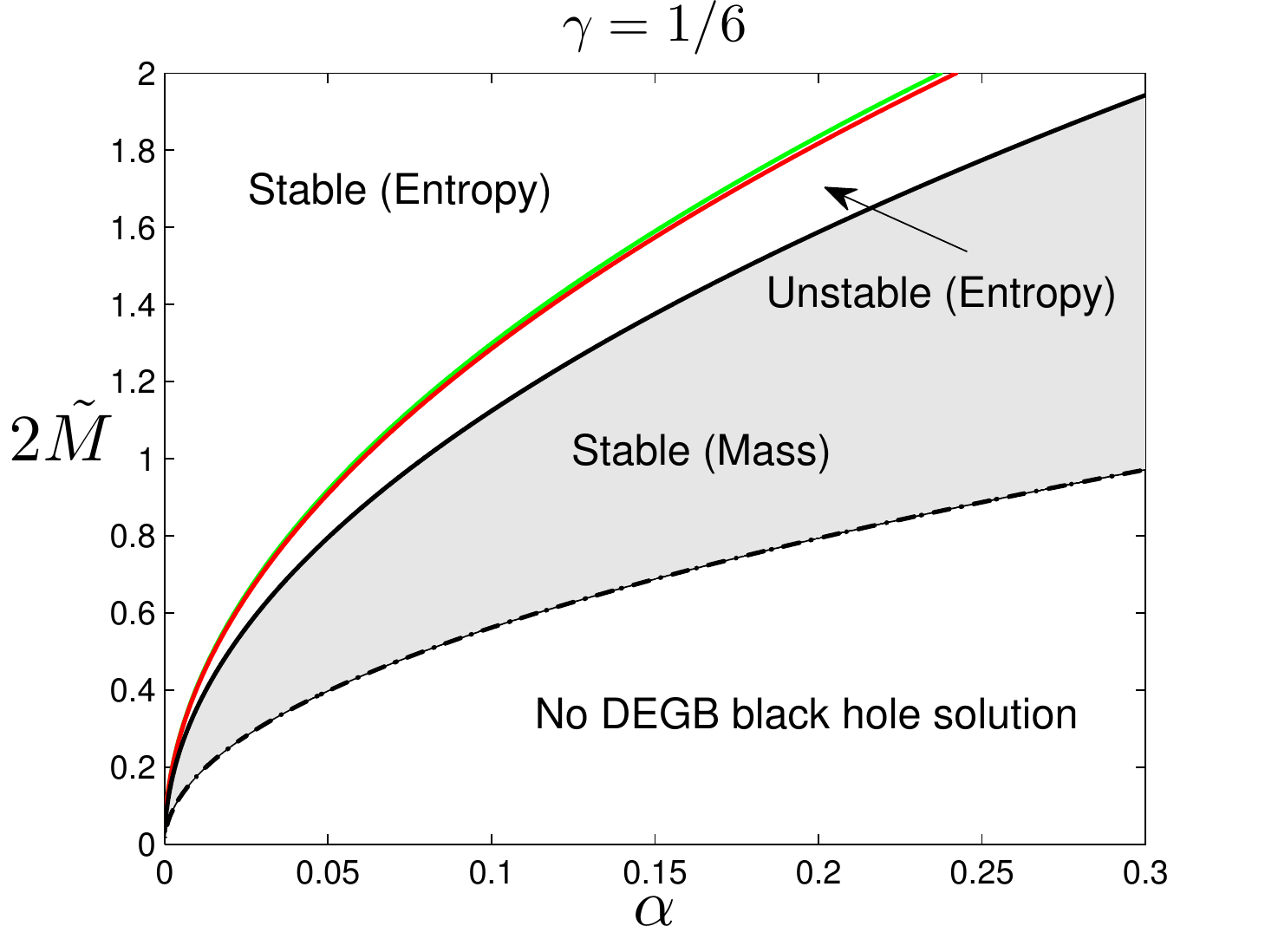}}
\end{center}
\vspace{-0.6cm}
\caption{\footnotesize{The phase diagrams with respect to $\alpha$ and $\tilde{M}$ for fixed $\gamma$. The red solid line represents $(\frac{1}{2},\frac{1}{2})$} fragmentation. The green solid line represents $(\bar{\delta},1-\bar{\delta})$ fragmentation.}\label{fig:alphatm}
\end{figure}
In Fig.~\ref{fig:alphatm} for the given couplings $\gamma=\frac{1}{2}$ and $\gamma=\frac{1}{6}$, the DEGB black hole phases are represented with respect to mass $\tilde{M}$. The black hole mass $\tilde{M}$ is proportional to $\sqrt{\alpha}$, so the phase boundary is also proportional to $\sqrt{\alpha}$.  In the case of a large $\gamma$, as we see in Fig.~\ref{fig:estimate}(a), the DEGB black holes have three phases such as no DEGB black hole solution, and stable(mass) and stable(entropy) regions, as shown in Fig.~\ref{fig:alphatm}(a). In the case of a small coupling $\gamma$ as shown Fig.~\ref{fig:estimate}(b), the DEGB black holes have four phases such as no DEGB black hole solution, and stable(mass), unstable(entropy), and stable(entropy) regions, as shown in Fig.~\ref{fig:alphatm}(b). In the limit of $\gamma\rightarrow 0$, DEGB theory approaches the EGB theory, so the final phases are dominant for a small mass, and the black holes are unstable, as shown in Fig.~\ref{fig:alphatm}(b). The unstable region from fragmentation appears between the stable and absolutely stable region. The initial phase still is stable for a large mass. The largest unstable region comes from $(\bar{\delta},1-\bar{\delta})$ fragmentation. These unstable regions start at the origin of Fig.~\ref{fig:alphatm}(b). In the limit of $\alpha\rightarrow 0$, the stable(entropy) region covers all values of mass $2\tilde{M}$ as shown in Fig.~\ref{fig:alphatm}(b). The other regions such as the one having no DEGB black hole solution, and stable(mass), and unstable(entropy) regions disappear in the limit of $\alpha\rightarrow 0$. In other words, only the stable(entropy) region occurs, and the other regions have vanished at $\alpha\rightarrow 0$.

\begin{figure}[h]
\begin{center}
{\includegraphics[width=4.0in]{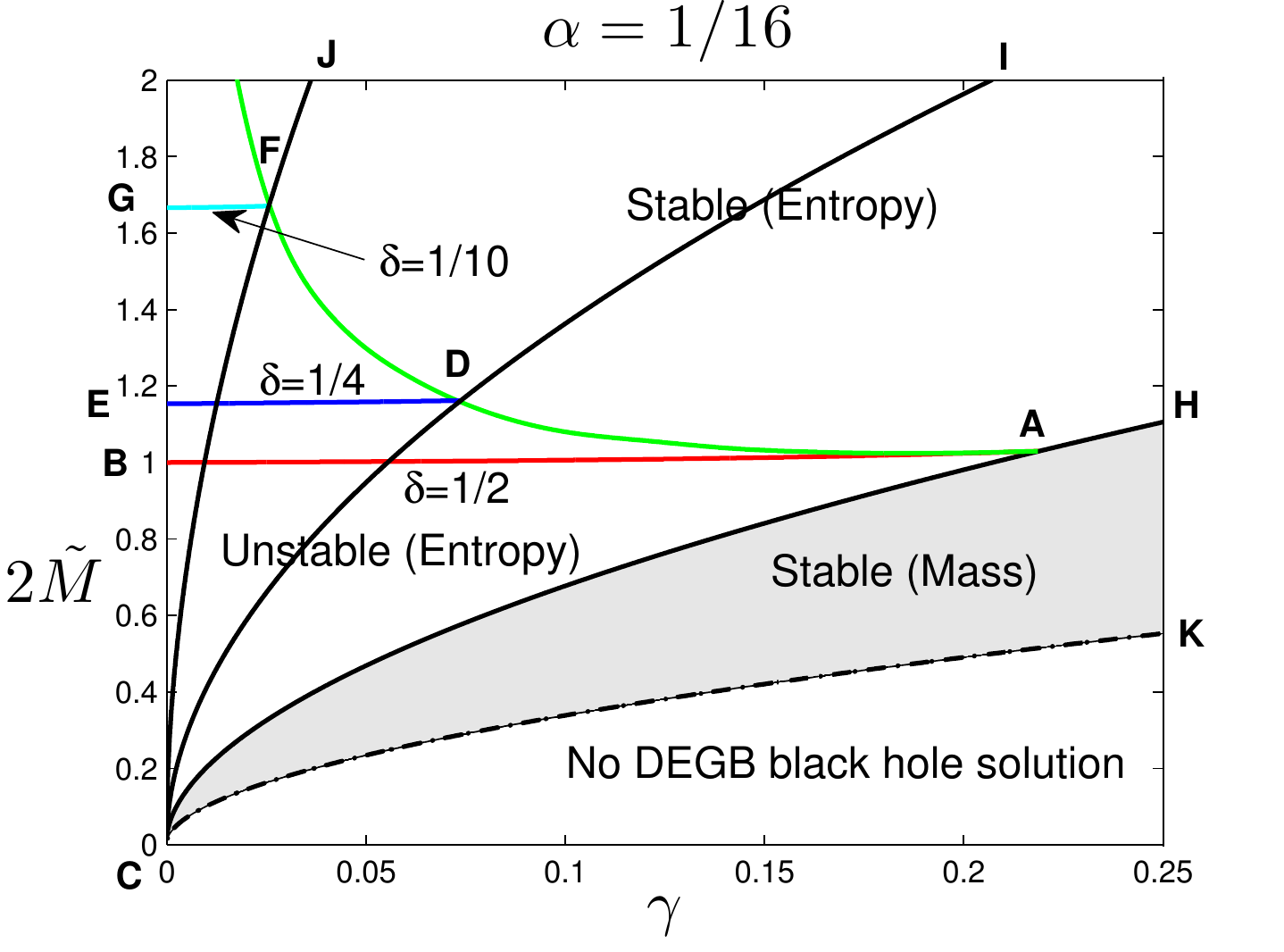}}
\vspace{-0.3cm}
\caption{\footnotesize{The phase diagrams with respect to $\gamma$ and $\tilde{M}$ in fixed $\alpha$ for $(\frac{1}{2},\frac{1}{2})$(red solid line), $(\frac{1}{4},\frac{3}{4})$(blue solid line), $(\frac{1}{10},\frac{9}{10})$(cyan solid line) and $({\bar{\delta}}$, $1-\bar{\delta})$(green solid line) fragmentation.}}\label{fig:hairymetric3}
\end{center}
\end{figure}
DEGB black hole phase diagrams are represented with respect to mass $\tilde{M}$ and $\gamma$ in Fig.~\ref{fig:hairymetric3}. DEGB black holes have stable(entropy), stable(mass), and no DEGB black hole solution phases for the large $\gamma$, while DEGB black holes have four phases such as stable(entropy), unstable(entropy), stable(mass), and no DEGB black hole solution phases for the small $\gamma$ as shown in Fig.~\ref{fig:hairymetric3}. The stable(mass) region is bounded by the minimum mass of the black hole. The $({\bar{\delta}}$, $1-\bar{\delta})$ fragmentation gives the largest unstable(entropy) region of the DEGB black holes and meets the stable(mass) region at twice the minimum mass or at $(\frac{1}{2},\frac{1}{2})$ fragmentation. The DEGB black hole is in the unstable(entropy) region for fixed $({\bar{\delta}}$, $1-\bar{\delta})$ horizontal blue line, as shown in Fig.~\ref{fig:hairymetric3}. For example, the stable(mass) region is larger in the case of $\delta=\frac{1}{4}$ fragmentation, such as the region $ICK$. The unstable(entropy) region becomes smaller, such as the region $ECD$. The stable(entropy) region is above the line $EDI$. The $({\bar{\delta}}$, $1-\bar{\delta})$ fragmentation is the marginal boundary of an arbitrary mass ratio fragmentation, so all of the fragmentations have ended at the $({\bar{\delta}}$, $1-\bar{\delta})$ fragmentation line, like $(\frac{1}{2},\frac{1}{2})$ fragmentation, as shown in Fig.~\ref{fig:hairymetric3}. For example, the fragmentations for $(\frac{1}{2},\frac{1}{2})$, $(\frac{1}{4},\frac{3}{4})$, and $(\frac{1}{10},\frac{9}{10})$ are shown in Fig.~\ref{fig:hairymetric3}. In the limit of $\gamma\rightarrow 0$, these behaviors correspond to the EGB black hole cases, so black hole solutions exist for all values of mass and have a crossing point in the $\gamma=0$ slice. In this limit, black holes only have two phases, the unstable(entropy) and stable(entropy) phases.

\section{Summary and discussion \label{sec5}}
We have investigated the fragmentation instability of black holes with a GB term. To explore these phenomena, we have numerically constructed the static DEGB hairy black hole in asymptotically flat spacetime. The two couplings $\alpha$ and $\gamma$ affect the scalar hair and mass of the black hole. The profiles of the scalar fields monotonically go to zero at the asymptotic region. The initial magnitudes of dilaton fields are almost inversely proportional to the black hole horizons. When the scalar field on the horizon is maximum, the DEGB black hole solution has a minimum horizon size, as shown in Fig.~\ref{fig:hairymetric}\,. The black hole solutions with respect to the coupling $\gamma$ are shown in Fig.~\ref{fig:tmr}\,. The black hole mass $\tilde{M}$ and horizon $\tilde{r}_h$ are proportional to $\sqrt{\alpha}$, so the black hole properties can be scaled with respect to $\alpha$ except $\alpha=0$, the Schwarzschild black hole case. The DEGB black hole has a minimum mass for given couplings. The amount of black hole hair decreases as the DEGB black hole mass increases. DEGB black hole configurations go to the EGB black hole cases for small $\gamma$. In the EGB black hole cases, the black hole solution exists for all values of the black hole mass. In other words, the minimum mass becomes zero.

We have investigated the DEGB black hole instability with fragmentation, which is based on thermal or quantum fluctuations. We found the unstable DEGB black hole phase under fragmentation, even if these phases are stable under perturbation. These instabilities have been numerically investigated with respect to the couplings.

In the limit of $\gamma \rightarrow 0$, the DEGB black hole approaches the EGB black hole. The EGB black hole simply has only two phases, the stable and unstable phases, under fragmentation. The small EGB black hole is unstable and is fragmented to a final phase. The relatively massive EGB black hole is stable. The mass ratio $\delta=\frac{1}{2}$ gives the smallest unstable region. In the limit of $\gamma \rightarrow 0$, EGB black holes are unstable. For the finite values of $\gamma$, the DEGB black hole has a minimum mass, so more phases appear. The mass ratio $\delta$ is bounded below $\bar{\delta}$. The mass ratio has a range between $\bar{\delta}$ and $\delta=\frac{1}{2}$. The phase diagrams for a given coupling are shown in Fig.~\ref{fig:alphatm}. For small $\gamma$, the DEGB black hole has four phases, such as the solution with no DEGB black hole, and the stable(mass), unstable(entropy), and stable(entropy) phases. There is no fragmented black hole solution between this minimum mass and twice the minimum mass, so the initial black hole is in the stable(mass) region with a mass in such a range. Above twice the minimum mass, the black hole can be fragmented with the mass ratio $(\delta,1-\delta)$. The fragmentation is bounded to $\bar{\delta}$ which is the minimum fragmentation for the given couplings. Above $\bar{\delta}$ fragmentation with respected to the black hole mass, the DEGB black hole gets in the stable(entropy) region under fragmentation. These phases reduce to three phases for large $\gamma$. The unstable(entropy) region under fragmentation approaches the stable(mass) region and then disappears. Above the minimum mass, the DEGB black hole is stable. In the limit of $\alpha$ to zero, the stable(entropy) region is dominant, and the other regions have disappeared.

The DEGB black hole phases are also shown in Fig.~\ref{fig:hairymetric3}. Through these diagrams, we can show that the $\bar{\delta}$ fragmentation plays the role of a marginal fragmentation. For given $\alpha$, the DEGB black hole has four phases, such as a solution with no DEGB black hole, and stable(mass), unstable(entropy), and stable(entropy) phases for small $\gamma$. For large $\gamma$, the DEGB black hole has three phases, such as solution with no DEGB black hole, and stable(mass), and stable(entropy) phases. These behaviors have not changed with respect to $\alpha$. The smallest unstable region comes from $\frac{1}{2}$ fragmentation, which meets at $\bar{\delta}$ fragmentation and the stable(mass) region. The mass ratio $\bar{\delta}$ fragmentation gives the largest unstable(entropy) region. The $\bar{\delta}$ fragmentation is the marginal fragmentation for any mass ratio. We have found the phase diagram of the fragmentation instability for a black hole mass and two couplings.

\section{Acknowledgements}

We would like to thank Nobuyoshi Ohta, Yun Soo Myung, Jin Young Kim, Gungwon Kang, and Sunly Khimphun for helpful discussions and comments. This work was supported by the National Research Foundation of Korea(NRF) grant funded by the Korea government(MSIP) (No.~2014R1A2A1A010). BG was supported by Basic Science Research Program through the National Research Foundation of Korea(NRF) funded by the Ministry of  Science, ICT \& Future Planning(2015R1C1A1A02037523). WL was supported by Basic Science Research Program through the National Research Foundation of Korea(NRF) funded by the Ministry of Education, Science and Technology(2012R1A1A2043908). We acknowledge the hospitality at KIAS where part of this work was done.

\newpage

{\bf Appendix A}

{\small
The detailed forms of the functions in Eq.~(\ref{twoeq}) are as follows:
\begin{eqnarray}
W&=&4 \left[8 \left(-1+e^Y\right) \alpha  \gamma  X' \left\{-3 e^{2 (Y+\gamma  \Phi )} r^2+4 e^{Y+\gamma  \Phi } \left(-13+3 e^Y\right) r \alpha \gamma  \kappa  \Phi ' \right. \right. \nonumber\\
&&\left.+16 \left(-15+7 e^Y\right) \alpha ^2 \gamma ^2 \kappa ^2 {\Phi '}^2\right\}+ e^{Y+\gamma  \Phi } \left\{24 e^{Y+\gamma  \Phi } \left(-1+e^Y\right)^2 r \alpha  \gamma  \right. \nonumber\\
&&\left.\left. +\left(-e^{2 (Y+\gamma  \Phi )} r^4+224 \alpha^2 \gamma ^2 \kappa -448 e^Y \alpha ^2 \gamma ^2 \kappa +224 e^{2 Y} \alpha ^2 \gamma ^2 \kappa \right) \Phi '  \right. \right. \nonumber\\
&&\left.\left. +4 e^{Y+\gamma  \Phi } \left(-7+3 e^Y\right) r^3 \alpha  \gamma  \kappa  \left(\Phi '\right)^2+32 \left(-5+3 e^Y\right) r^2 \alpha ^2 \gamma ^2 \kappa ^2 \left(\Phi '\right)^3\right\}\right], \nonumber\\
W1&=&2\left(X'\right)^2 \left(e^{Y+\gamma  \Phi } r+8 \alpha  \gamma  \kappa  \Phi '\right) \left\{e^{Y+\gamma  \Phi } r-4
\left(-3+e^Y\right) \alpha  \gamma  \kappa  \Phi '\right\}^2 \nonumber\\
&&+X' \left\{10 e^{3 (Y+\gamma  \Phi )} \left(-1+e^Y\right) r^2-4 e^{2 (Y+\gamma  \Phi )} r \gamma  \left(e^{Y+\gamma  \Phi } r^2+42 \alpha  \kappa-52 e^Y \alpha  \kappa \right.\right. \nonumber\\
&&\left.+10 e^{2 Y} \alpha  \kappa \right) \Phi '- e^{Y+\gamma  \Phi } \kappa  \left(5 e^{2 (Y+\gamma  \Phi )} r^4+32 e^{Y+\gamma  \Phi } r^2 \alpha  \gamma ^2+64 e^{2 Y+\gamma  \Phi } r^2 \alpha \gamma ^2 \right. \nonumber\\
&&\left. +768 \alpha ^2 \gamma ^2 \kappa -1152 e^Y \alpha ^2 \gamma ^2 \kappa +384 e^{2 Y} \alpha ^2 \gamma ^2 \kappa \right) \left(\Phi '\right)^2+4 e^{Y+\gamma  \Phi } r \alpha  \gamma  \left(-23 e^{Y+\gamma  \Phi } r^2 \right. \nonumber\\
&&\left.+5 e^{2 Y+\gamma  \Phi } r^2+80 \alpha  \gamma ^2-352 e^Y \alpha\gamma ^2+80 e^{2 Y} \alpha  \gamma ^2\right) \kappa ^2 \left(\Phi '\right)^3+32 \alpha ^2 \gamma ^2 \left(-15 e^{Y+\gamma  \Phi } r^2 \right. \nonumber\\
&&\left.\left.+7 e^{2 Y+\gamma \Phi } r^2+96 \alpha  \gamma ^2-256 e^Y \alpha  \gamma ^2+96 e^{2 Y} \alpha  \gamma ^2\right) \kappa ^3 \left(\Phi '\right)^4\right\}\nonumber\\
&&+2 e^{Y+\gamma  \Phi } \left\{-6 e^{2 (Y+\gamma  \Phi )} \left(-1   +e^Y \right)^2 r+2 e^{Y+\gamma  \Phi } \left(-1+e^Y\right) \gamma  \left(e^{Y+\gamma \Phi } r^2  \right.\right. \nonumber\\
&&\left.+28 \alpha  \kappa -28 e^Y \alpha  \kappa \right) \Phi '+e^{Y+\gamma  \Phi } r \left(3 e^{Y+\gamma  \Phi } r^2+e^{2 Y+\gamma  \Phi } r^2+8 \alpha  \gamma ^2-48 e^Y \alpha  \gamma ^2 \right. \nonumber\\
&&\left. +40 e^{2 Y} \alpha  \gamma ^2\right) \kappa  \left(\Phi '\right)^2+\gamma  \kappa  \left(e^{2 (Y+\gamma  \Phi )} r^4+76 e^{Y+\gamma  \Phi } r^2 \alpha  \kappa -12 e^{2 Y+\gamma  \Phi } r^2 \alpha  \kappa \right. \nonumber\\
&&\left. +256\alpha ^2 \gamma ^2 \kappa -640 e^Y \alpha ^2 \gamma ^2 \kappa +384 e^{2 Y} \alpha ^2 \gamma ^2 \kappa \right) \left(\Phi '\right)^3+r \kappa ^2 (e^{2 (Y+\gamma  \Phi )} r^4 \nonumber\\
&&+4 e^{Y+\gamma  \Phi } r^2 \alpha  \gamma ^2+12 e^{2 Y+\gamma  \Phi } r^2 \alpha  \gamma^2+480 \alpha ^2 \gamma ^2 \kappa -256 e^Y \alpha ^2 \gamma ^2 \kappa \nonumber\\
&&\left.\left.+32 e^{2 Y} \alpha ^2 \gamma ^2 \kappa \right) \left(\Phi '\right)^4 +8 r^2\alpha  \gamma  \left(e^{Y+\gamma  \Phi } r^2-4 \alpha  \gamma ^2+12 e^Y \alpha  \gamma ^2\right) \kappa ^3 \left(\Phi '\right)^5 \right\}, \nonumber\\
W2&=&-8 \alpha  \gamma  \left(X'\right)^3 \left\{-e^{2 (Y+\gamma  \Phi )} \left(-7+5 e^Y\right) r^2+4 e^{Y+\gamma  \Phi } \left(29-26
e^Y+5 e^{2 Y}\right) r \alpha  \gamma  \kappa  \Phi ' \right. \nonumber\\
&&\left. +32 \left(15-16 e^Y +5 e^{2 Y}\right) \alpha ^2 \gamma ^2 \kappa ^2 \left(\Phi '\right)^2\right\}+2 e^{2 (Y+\gamma  \Phi )} \Phi ' \left\{2 \left(-1+e^Y\right) \left(e^{Y+\gamma  \Phi } r^2 \right. \right. \nonumber\\
&&\left.\left.+8 \alpha  \gamma ^2-8 e^Y \alpha  \gamma ^2\right)-16\left(1-4 e^Y+3 e^{2 Y}\right) r \alpha  \gamma  \kappa  \Phi '+r^2 \left(e^{Y+\gamma  \Phi } r^2+24 \alpha  \gamma ^2 \right.\right.\nonumber\\
&&\left.\left.-24 e^Y \alpha  \gamma ^2\right)\kappa  \left(\Phi '\right)^2-8 \left(-5+3 e^Y\right) r^3 \alpha  \gamma  \kappa ^2 \left(\Phi '\right)^3\right\}+2 e^Y \left(X'\right)^2 \left\{-20 e^{Y+2 \gamma  \Phi } \left(-1  \right. \right. \nonumber\\
&&\left. +e^Y\right)^2 r \alpha  \gamma +e^{\gamma  \Phi } \left(e^{2 (Y+\gamma  \Phi)} r^4+8 e^{Y+\gamma  \Phi } r^2 \alpha  \gamma ^2+8 e^{2 Y+\gamma  \Phi } r^2 \alpha  \gamma ^2-128 \alpha ^2 \gamma ^2 \kappa \right.\nonumber\\
&&\left.+256 e^Y \alpha ^2\gamma ^2 \kappa -128 e^{2 Y} \alpha ^2 \gamma ^2 \kappa \right) \Phi '-2 e^{\gamma  \Phi } r \alpha  \gamma  \left(-25 e^{Y+\gamma  \Phi } r^2+7 e^{2 Y+\gamma  \Phi } r^2-48 \alpha  \gamma ^2 \right. \nonumber\\
&&\left.-96 e^Y \alpha \gamma ^2 +16 e^{2 Y} \alpha  \gamma ^2\right) \kappa  \left(\Phi '\right)^2-16 \alpha ^2 \gamma ^2 \left(-25 e^{\gamma  \Phi } r^2+13 e^{Y+\gamma \Phi } r^2-96 \alpha  \gamma ^2  \right. \nonumber\\
&&\left.\left.+32 e^Y \alpha  \gamma ^2\right) \kappa ^2 \left(\Phi '\right)^3\right\}+e^{Y+\gamma  \Phi } X' \left\{16 e^{Y+\gamma  \Phi } \left(-1+e^Y\right)^2 \alpha  \gamma +2 e^{Y+\gamma  \Phi } r \left(e^{Y+\gamma  \Phi }r^2 \right.\right.\nonumber\\
&&\left.+e^{2 Y+\gamma  \Phi } r^2-8 \alpha  \gamma ^2+16 e^Y \alpha  \gamma ^2-8 e^{2 Y} \alpha  \gamma ^2\right) \Phi '-8 \alpha  \gamma  \left(9 e^{Y+\gamma \Phi } r^2-5 e^{2 Y+\gamma  \Phi } r^2 \right.\nonumber\\
&&\left.+48 \alpha  \gamma ^2-96 e^Y \alpha  \gamma ^2+48 e^{2 Y} \alpha  \gamma ^2\right) \kappa  \left(\Phi '\right)^2-r \kappa  \left(e^{2 (Y+\gamma  \Phi )} r^4+24 e^{Y+\gamma  \Phi } r^2 \alpha  \gamma ^2 \right.\nonumber\\
&&\left.+24 e^{2 Y+\gamma  \Phi } r^2 \alpha  \gamma ^2+960\alpha ^2 \gamma ^2 \kappa -768 e^Y \alpha ^2 \gamma ^2 \kappa +320 e^{2 Y} \alpha ^2 \gamma ^2 \kappa \right) \left(\Phi '\right)^3 \nonumber\\
&&\left.-8 e^Y r^2 \alpha \gamma  \left(5 e^{\gamma  \Phi } r^2+48 \alpha  \gamma ^2\right) \kappa ^2 \left(\Phi '\right)^4\right\}. \nonumber
\end{eqnarray}
}

\newpage

{\bf Appendix B}

{\small
The initial conditions for $\Phi_h$ and $r_h$ satisfy the inequality in Eq.~(\ref{bcphi}) for the given couplings $\alpha$ and $\gamma$. The different choices of $r_h$ provide different values of $\Phi_h$ as shown in Fig.~\ref{fig:apb1}(a). The minimum values of $\Phi_h$ for each solid line satisfy the inequality in Eq.~(\ref{bcphi}). Each solid line gives different profiles $\Phi(r)$\,. As a result, the values of the scalar fields at infinity $\Phi_{\infty}$ are different for each solid line, as shown in Fig.~\ref{fig:apb1}(b).
\begin{figure}[ht]
\begin{center}
\subfigure []{\includegraphics[width=3.0in]{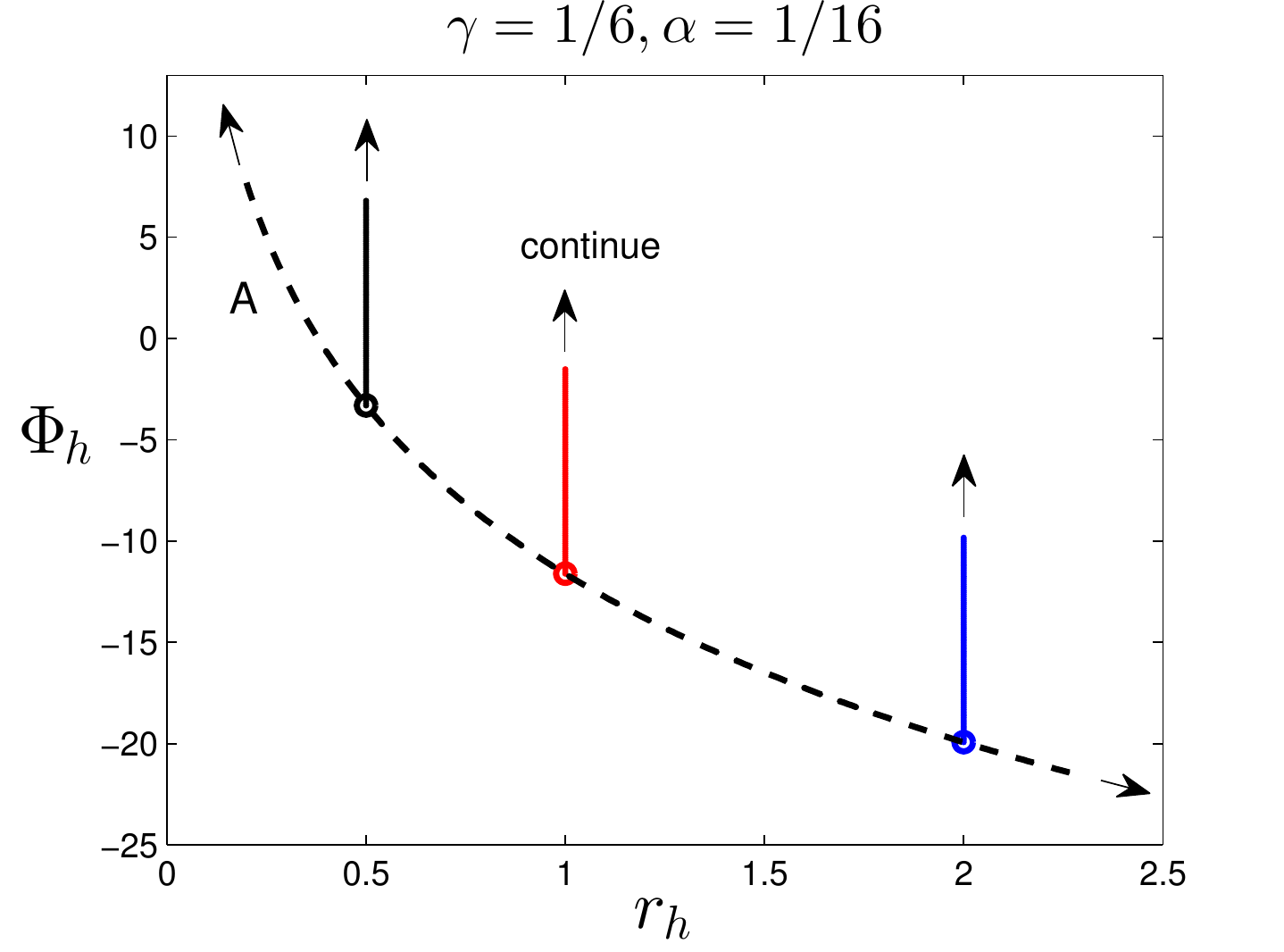}}
\subfigure []{\includegraphics[width=3.0in]{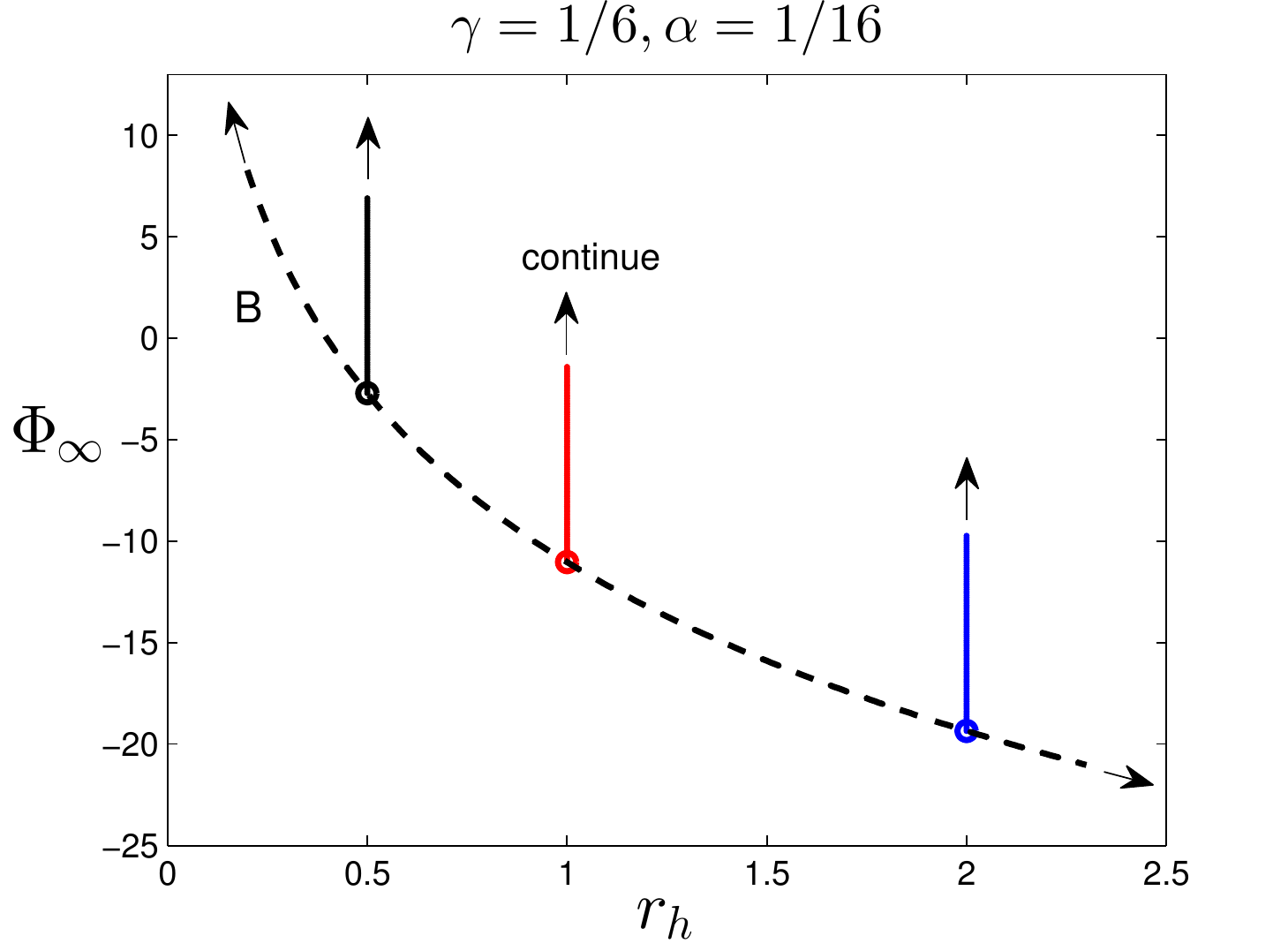}}
\end{center}
\vspace{-0.6cm}
\caption{\footnotesize{(a) The allowed values of $\Phi_h$ for given $r_h=1/2$, $1$, and $2$ with given couplings. (b) The values of $\Phi(r)$ for $r \rightarrow \infty$ for given $r_h=1/2$, $1$, and $2$ with given couplings.}}\label{fig:apb1}
\end{figure}
Using scalar field values $\Phi_\infty$, we obtain a rescaled system, as shown in Fig.~\ref{fig:apb2}. For each $r_h$ choice, the rescaled scalar fields $\tilde{\Phi}_h$ are rearranged in Fig.~\ref{fig:apb2}(a), and the rescaled scalar field values $\tilde{\Phi}_h$ are all the same for different choices of $r_h$.
\begin{figure}[ht]
\begin{center}
\subfigure []{\includegraphics[width=3.0in]{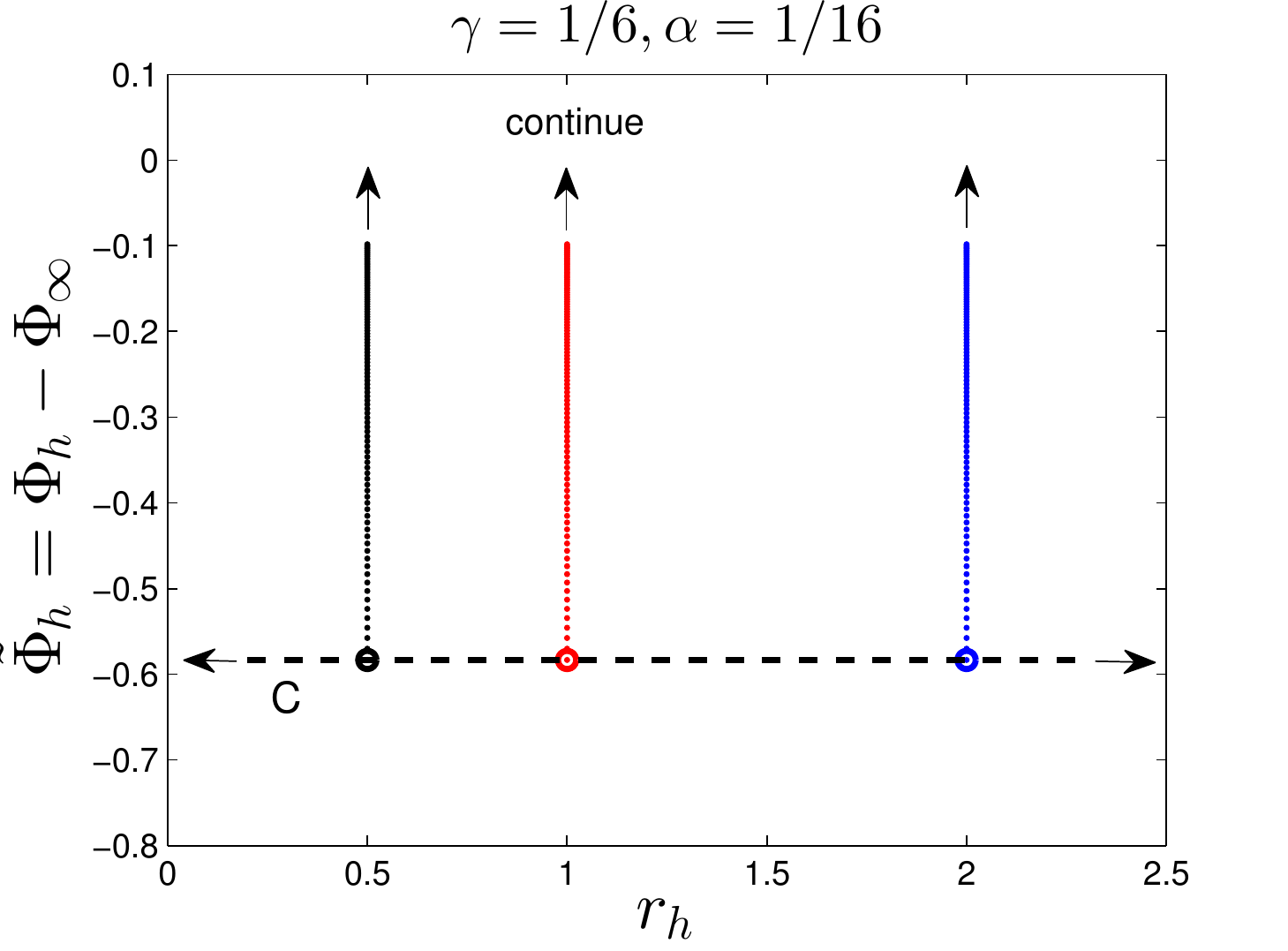}}
\subfigure []{\includegraphics[width=3.0in]{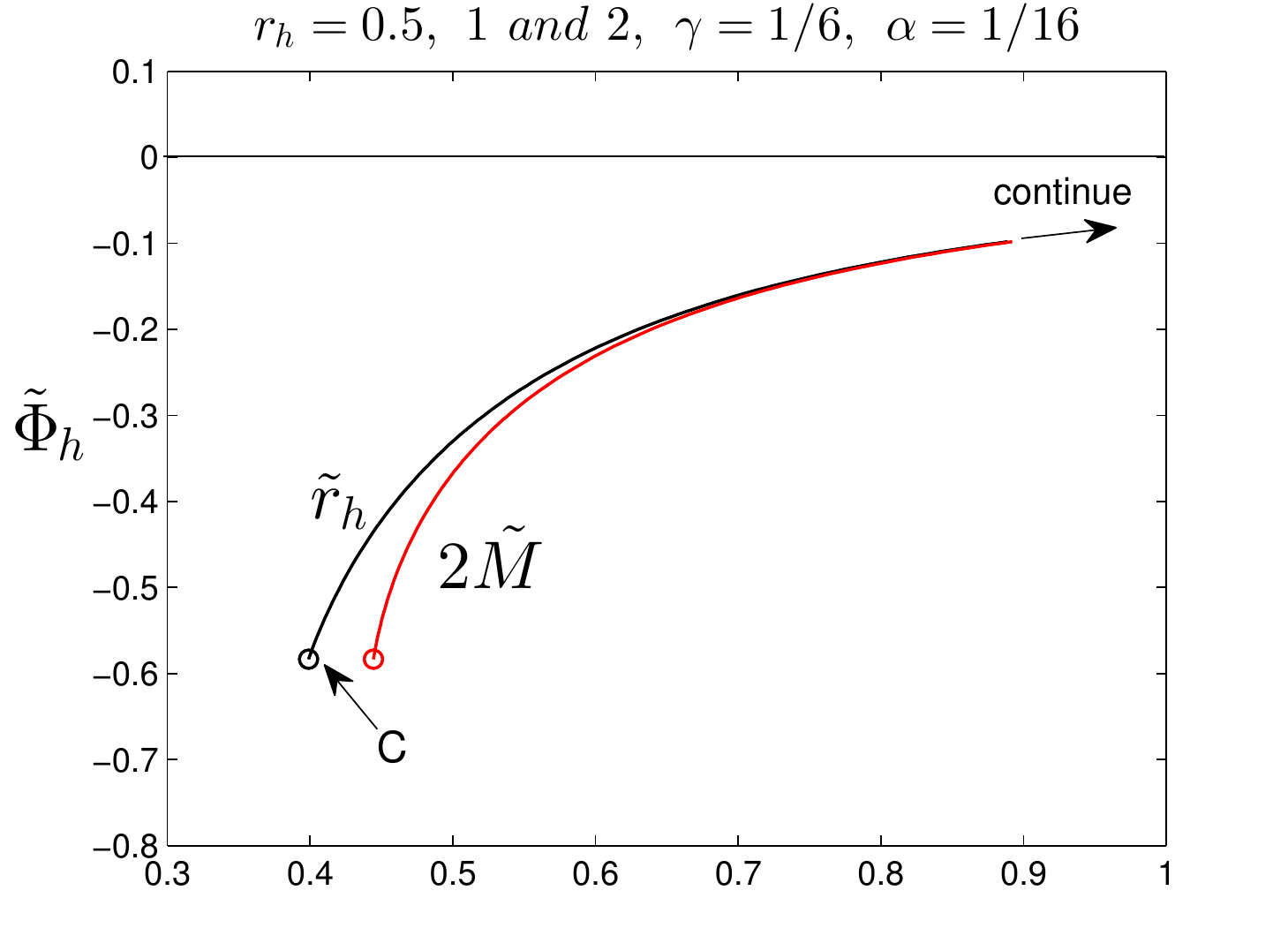}}
\end{center}
\vspace{-0.6cm}
\caption{\footnotesize{(a) The initial conditions $\tilde{\Phi}_h$ for given $r_h=1/2$, $1$, and $2$ with given couplings. (b) The different initial conditions of given $r_h=1/2$, $1$, and $2$ converge to a line with respect to the black hole mass $\tilde{M}$ or horizon $\tilde{r}_h$.}}\label{fig:apb2}
\end{figure}
Next, the horizon radii are also rescaled to $\tilde{r}_h$. Eventually, the different choices of $\Phi_h$ and $r_h$ converge to a unique black sold line in Fig.~\ref{fig:apb2}(b). In other words, the 2-dimensional solution space reduces to an actually 1-dimensional line. Therefore, whatever we choose for any value of $r_h$, there is no loss of generality.

\end{document}